\documentclass[pdflatex]{article}
\usepackage{mathptmx}
\usepackage{amsmath}
\usepackage{amssymb}
\usepackage{manyfoot}
\usepackage{helvet}

\usepackage[LGR,T1]{fontenc}
\usepackage[latin9]{inputenc}
\usepackage{geometry}
\usepackage{babel}
\usepackage{array}
\usepackage{multirow}
\usepackage{colortbl}
\usepackage{amsbsy}
\usepackage{float}
\usepackage{graphicx}
\usepackage{subscript}
\usepackage{subfig}
\usepackage{tablefootnote}

\usepackage[bookmarks=false]{hyperref}
\hypersetup{colorlinks, linkcolor=blue,  citecolor=blue, urlcolor=blue}

\DeclareRobustCommand{\greektext}{%
  \fontencoding{LGR}\selectfont\def\encodingdefault{LGR}}
\DeclareRobustCommand{\textgreek}[1]{\leavevmode{\greektext #1}}
\ProvideTextCommand{\~}{LGR}[1]{\char126#1}

\providecommand{\tabularnewline}{\\}
\floatstyle{ruled}
\newfloat{algorithm}{tbp}{loa}
\providecommand{\algorithmname}{Algorithm}
\floatname{algorithm}{\protect\algorithmname}

\begin{document}

\title{Relay-Based Synchronization of Replicated Data Types\\ in Opportunistic Networks}

\date{~}

\author{
 Fr\'ed\'eric Guidec \\ IRISA, Universit\'e Bretagne Sud, France \\ Frederic.Guidec@univ-ubs.fr
\and
Yves Mah\'eo \\ IRISA, Universit\'e Bretagne Sud, France \\Yves.Maheo@univ-ubs.fr
}

\maketitle

\begin{abstract}
In Opportunistic Networks (OppNets), the dissemination of information can only
rely on transient pairwise radio contacts between mobile devices
(peers). Designing distributed applications that can run in such conditions is
a challenge, but replicated data types, and in particular Conflict-free
Replicated Data Types (CRDTs), can help meet this challenge. A CRDT is
inherently a replicated data type whose replicas can be updated locally, yet
eventually converge thanks to an anti-entropy algorithm that allows all
replicas to synchronize in the background.  Whether the replicas of a CRDT can
actually converge in an OppNet, and how fast they can converge, depend on the
occurrence of radio contacts between mobile devices. In this paper we
investigate the idea of using mobile relays as a means to boost the convergence
of stated-based CRDT replicas in an OppNet. New protocols are presented that allow the
synchronization of replicas and relays, and new metrics are defined to observe
and characterize the convergence of replicas. Simulation results show that
using relays can significantly improve this convergence, and even make it
possible in scenarios where the replicas alone would be unable to converge.\\

\noindent\textbf{Keywords:} Opportunistic networking, Peer-to-Peer networks, Ad hoc networks, Conflict-free replicated data types, Optimistic replication, Strong eventual consistency
\end{abstract}

~\\

\section{\label{sec:introduction}Introduction}

\subsection{Opportunistic networking}

An opportunistic network (OppNet) is a category of peer-to-peer network
in which the peers are mobile devices that can only interact pairwise
via direct radio transmission links. Such networks can typically be
used when no communication infrastructure is available (e.g., in disaster
relief scenarios, or in V2V communication), or as a means to complement
an existing infrastructure.

\begin{figure*}[h]
\begin{centering}
\includegraphics[width=15cm]{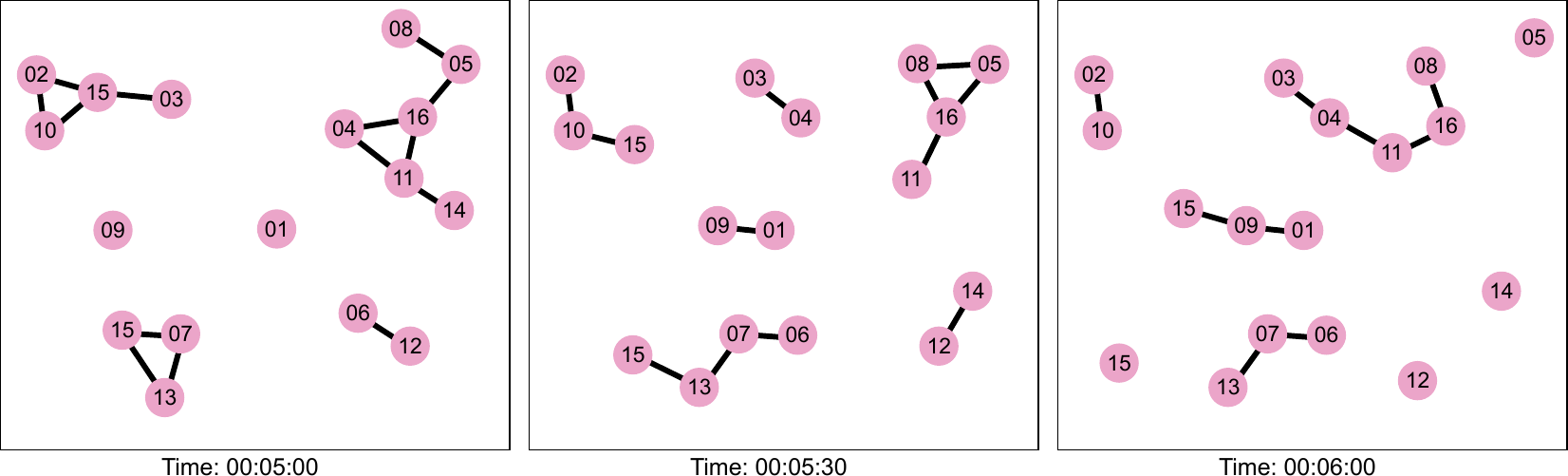}
\par\end{centering}
\centering{}\caption{\label{fig:snapshots_oppnet}Illustration of the evolution of
an opportunistic network}
\end{figure*}

Since an OppNet is often sparsely populated, and since the radio links
between neighbor devices (or peers) are often short ranged, there
is usually very little connectivity between the devices. An OppNet,
if \foreignlanguage{american}{modelled} as a graph, appears as a highly
partitioned time-varying graph rather than as a connected static graph
(see Figure~\ref{fig:snapshots_oppnet}). A consequence of the
 high level of partitioning in such a network is that routing algorithms
such as those used in the Internet, or even in MANETs, are useless
in OppNets. While permanent or quasi-permanent end-to-end connectivity
between any pair of hosts is taken for granted in the Internet, it
is simply unavailable in OppNets. For this reason, maintaining an
overlay among peers is not an option. Each mobile device in an OppNet
can only seize the \emph{opportunity} to interact transiently with
another mobile device (or peer) when a radio contact is established
between them.

The very nature of OppNets requires that applications be fully distributed
rather than based on the client-server architecture. Yet the development
of distributed applications capable of running in OppNets, relying
solely on transient radio contacts between neighbor nodes, is quite
a challenge. 

\subsection{Conflict-free Replicated Data Types}

Conflict-free Replicated Data Types (CRDTs) can significantly help
in the development of distributed applications in OppNets because
of their ability to provide Strong Eventual Consistency (SEC) for
replicated data types~\cite{computingsurveys24almeida,inria11shapiro}.
They can typically be used to implement distributed databases, as
well as collaborative editing applications. A CRDT is basically a
data type that can be replicated on many hosts. An update operation
can be issued locally on any replica at any time, regardless of what
may be happening on other replicas at the same time. A synchronization
protocol (also called anti-entropy algorithm) runs in the background to ensure
that each update is eventually delivered to ---and applied by---
every replica (eventual visibility/delivery), and replicas that have
applied the same updates have equivalent states (strong convergence).
With this consistency model, replicas keep converging as long as new
updates are issued. If no new update is issued for long enough the
replicas eventually reach equivalent states.

CRDTs come in two main flavors: op-based CRDTs (i.e., operation-based CRDT,
sometimes called commutative replicated data types), and state-based CRDTs
(also known as convergent replicated data types). It has been demonstrated that
op-based CRDTs can hardly be used in OppNets~\cite{ppna22guidec}, because their
synchronization model yields too much traffic in the network. In contrast,
state-based CRDTs are perfectly suited to be used in OppNets, as discussed
below.

In a state-based CRDT, the set of all possible states constitute a
join semi-lattice. A join operator is implemented in a merge function,
which takes two states as parameters and returns the least upper bound
of both states. Replicas can therefore synchronize periodically (usually
pairwise in a peer-to-peer manner) by exchanging their entire states,
each receiver merging the received state with its own local state,
so the two replicas involved end up with the same state. Causal consistency
is ensured implicitly, since successive synchronizations between pairs
of replicas ensure a transitive propagation of the causal past~\cite{computingsurveys24almeida}.

\begin{figure*}[h]
\begin{centering}
\includegraphics[width=\textwidth]{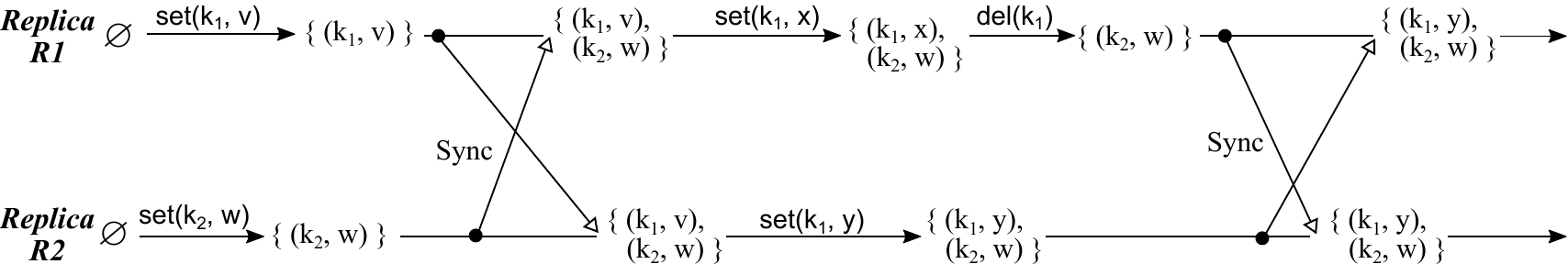}
\par\end{centering}
\vspace*{3mm}
\caption{\label{fig:map_crdt}Example of a run involving a Map CRDT (with set-wins semantics) replicated in two replicas R1 and R2}
\end{figure*}

An illustration of how replicas can periodically synchronize pairwise is
provided in Figure~\ref{fig:map_crdt}. In this example we consider two replicas
of a Map CRDT. To keep it simple, we assume that the values are of simple types
(integer, float, string...). This map is initially empty on both replicas. An
entry with key $k_{1}$ and value $v$ is first added locally to the map in
replica R1, while an entry $(k_{2},w)$ is added locally in replica R2. The
state is thus temporarily different in R1 and R2, but as a synchronization
occurs between them, R1 merges its local state with the state received from R2,
and R2 likewise merges its own state with that received from R1. After this
synchronisation, both replicas agree that the state of the map is now $\left\{
(k_{1},v),(k_{2},w)\right\} $.  Note that reconciling state $\left\{
(k_{1},v)\right\} $ (from R1) with state $\left\{ (k_{2},w)\right\} $ (from R2)
is not an issue, because these are not conflicting states: although
{\footnotesize \textsf{set(k$_{\mathsf 1}$, v)}} and {\footnotesize
  \textsf{set(k$_{\mathsf 2}$, w)}} occurred concurrently in R1 and R2, they
apply to different entries (i.e., entries with different keys) in the shared
map.

After the first synchronization, the value of entry $k_{1}$ is first
changed to $x$ in R1, and this entry is then deleted in R1, while
an entry with the same key $k_{1}$ and value $y$ is set in R2. The
state is thus different again in R1 and R2, and this time the last
{\footnotesize \textsf{del(k$_{\mathsf 1}$)}} on R1 conflicts
with {\footnotesize \textsf{set(k$_{\mathsf 1}$, y)}} 
on R2, as they occurred concurrently and both apply to entry $k_{1}$.
If both replicas synchronize again, the final state depends on the
concurrency semantics implemented in the merge function of the Map
CRDT. If this function implements the set-wins semantics,
giving {\footnotesize \textsf{set(k$_{\mathsf 1}$, x)}} priority
over {\footnotesize \textsf{del(k$_{\mathsf 1}$)}},
the final state is $\left\{ (k_{1},y),(k_{2},w)\right\} $
in both replicas, as shown in Fig.~\ref{fig:map_crdt}. But the
merge function could also implement the del-wins semantics, giving
{\footnotesize \textsf{del(k$_{\mathsf 1}$)}}  priority
over {\footnotesize \textsf{set(k$_{\mathsf 1}$, x)}}, in which
case the final state on both replicas would be $\left\{ (k_{2},w)\right\} $.
Both concurrency semantics are thus perfectly viable, as they allow
the replicas to reach the same state eventually. It is crucial that
all replicas of a state-based CRDT use the same merge function, and
thus the same concurrency semantics, when conflicting updates must
be dealt with. In practice, every state-based CRDT library implements
one or several builtin concurrency semantics for each kind of CRDT,
so an application developer simply needs to choose an implementation
that meets the application's needs.

An interesting characteristic of state-based CRDTs is that they do
not require reliable transmissions among replicas. Since the merge
function is commutative, associative, and idempotent, messages carrying
replicas' states may get lost, or get delivered several times and
in any order. Moreover, some hosts may crash and the network may get
partitioned episodically. Yet all replicas converge eventually, as
long as the synchronization graph is connected~\cite{arxiv18preguica,arxiv13almeida}. 

The ability of state-based CRDTs to tolerate partitions and unreliable
transmissions makes them ideal candidates to be used as software building
blocks in OppNets. In~\cite{ubicomm23maheo} it has been shown that
collaborative editing can be achieved in a small OppNet, using a CRDT-based
application for document editing among a few contributors. Yet, in a larger
OppNet, involving possibly hundreds or thousands of nodes, there is no reason
to presume that all the nodes should necessarily run the same distributed
application, hosting replicas of the same CRDT. In a large OppNet it is
actually likely that, for each CRDT instance, only a fraction of the network's
population would be interested in this particular CRDT, and would thus host a
replica of this CRDT. The lower the density of nodes carrying replicas of a
particular CRDT, the lower the chance for these nodes to get in radio contact,
and thus to synchronize their replicas.

\subsection{Using relays to assist state-based CRDT synchronization in Oppnets}

In this paper we investigate the novel idea of enrolling nodes as relays
to assist in the synchronization of replicas. We assume that all nodes
in an OppNet do not necessarily host replicas of a given CRDT instance,
so the nodes that actually host replicas struggle to synchronize these
replicas. Yet other nodes can assist in this synchronization by carrying
replicas' states (or more precisely, snapshots of these states), thus
bridging the gap between replicas and increasing the chance for these
replicas to synchronize.

Traditionally, applications running in OppNets rely on message routing
or message dissemination based on the store-carry-and-forward principle~\cite{sigcomm03fall}.
According to this principle, any mobile node can contribute to the
propagation of a message by carrying for a while this message, stored
in a cache, before it can be forwarded to another node (which itself
may be the destination of the message, or another mobile carrier for
this message).

In the present work we explore a similar idea and propose to improve the
synchronization of state-based CRDTs in OppNets by using relay nodes
to assist the replicas. Yet, unlike the above-mentioned message forwarding
protocols, which focus on message propagation via the store-carry-and-forward
principle, we focus on the propagation of replicas' states. This is
a major difference, for in message forwarding the payload of a message
is inherently an immutable object that is meant to reach its destination(s),
ideally, without being altered or dropped during its journey. In contrast,
a replica's state is a mutable object: it changes continuously, either
as update operations are issued on the replica, or by being merged
with other states. If a replica's state is serialized and the resulting
snapshot is transferred to a relay, this snapshot is not necessarily
meant to propagate throughout the network until it reaches a target
replica (or several target replicas, for that matter). It may become
rapidly obsolete and get superseded by more recent snapshots issued
either by the same source replica, or by another replica. The snapshot
of a replica's state is thus meant to propagate in the network only
as long as it cannot be superseded by another snapshot.

The protocols designed to support message forwarding in OppNets all
strive to deliver each message to its set destination. They are not
meant to account for messages whose payloads are causally ordered,
and can thus supersede each other. Using a message forwarding protocol
to propagate replicas' states in an OppNet would be utterly inefficient,
as the relays would inevitably store, carry and forward messages whose
content is not relevant anymore for the target replicas.

Thus, instead of propagating messages up to their destination, we
need to propagate states from node to node, ensuring that each intermediate
node (i.e., relay) only maintains a state ---and contributes to propagate
it further--- if this state is still relevant for the synchronization
process. 

\subsection{Paper contributions}

\paragraph{A Relay-Based Synchronization System}

The main contribution of this paper is the definition of a Relay-Based
Synchronization System (RBSS) that makes it easier to synchronize the
replicas of state-based CRDTs in an OppNet, thanks to the use of relay
nodes capable of carrying replica's states. This RBSS is meant to
fullfill the following objectives:

\begin{itemize}
\item Genericity: a relay should be able to carry replicas'
states, regardless of the type of CRDT considered (register, set,
map, list, etc.), and regardless of the CRDT library used to implement
the replicas (e.g., Yjs\footnote{\noindent \href{https://github.com/yjs/yjs}{https://github.com/yjs/yjs}},
Automerge\footnote{\href{https://www.automerge.org}{https://www.automerge.org}} or Loro\footnote{\href{https://www.loro.dev}{https://www.loro.dev}}). 

\item Confidentiality: although the application code
that uses CRDTs must be able to read the state of a replica, and to
modify this state by issuing update operations or merging it with
the state of another replica, it is neither necessary nor desirable
that a relay be able to read and modify the replicas' states it is
carrying. On the contrary, a relay should perceive each state issued
by a replica as a blob of opaque data (typically a byte array), which
may be encrypted and signed by the source replica so that only peer
replicas can check its validity and decrypt its content.

\item Frugality: the protocols used to ensure the
synchronization of a replica with a relay, and the synchronization of
two relays, should be designed so as to minimize the amount of data
transferred during a synchronization. The amount of data stored on a
relay should likewise be minimized. Both objectives aim at avoiding
that behaving as a relay yields too much overhead for the nodes
concerned, and too much traffic on the wireless channel.

\end{itemize}

\paragraph{Evaluation metrics}

In order to evaluate how much faster replicas can converge when relays
are used to assist in the synchronization, we define two novel metrics
called convergence latency and convergence distance, which are
especially relevant in OppNets. They show how a replica's state
deviates from an ideal reference state, that is, the single state that
would be observed if all updates were performed on all replicas with
no delay.

\paragraph{Experimentation results}

Simulations have been conducted in a variety of realistic scenarios in order to confirm that relays can significantly speed up the convergence of replicas. The results also confirm that relay-assisted synchronization is scalable up to scenarios involving hundreds of replicas or relays.

~\\
The remainder of this paper is organized as follows. Related work
is presented and discussed in Section~\ref{sec:Related_work}. The
system model we consider in this work is detailed in Section~\ref{sec:System_model}.
The protocols we designed to support replica-replica synchronization,
replica-relay synchronization, and relay-relay synchronization are
presented in Section~\ref{sec:Synchronization_algorithms}. Section~\ref{sec:Experimentation}
presents experimental results we obtained while running these protocols
in different scenarios. In Section~\ref{sec:Discussion} we address
additional questions, such as the scalability of the solution we propose
and the security issues ensuing from relay-based synchronization.
Section~\ref{sec:Conclusion} concludes the paper.

\section{\label{sec:Related_work}Related work}

Message routing (or message forwarding) has been an almost exclusive
focus in the research on OppNets in the last two decades, leading
to the definition of many protocols based on the store-carry-and-forward
principle~\cite{ieeeaccess22dalal}. These protocols enable point-to-point
transmission or some form of broadcast/multicast (e.g. with the publish/subscribe
paradigm). As a general rule, several copies of the original message
are allowed to propagate simultaneously in the network, in order to
maximize the chance to reach the destination(s), while minimizing
the time required to do so. Each mobile node that participates in
message routing serves as a mobile carrier: it is expected to store
messages in a local cache. While propagating in the network, a single
message can thus be stored simultaneously on many mobile carriers. 

Based on this general principle, many alternative forwarding
strategies have been proposed (pure epidemics, probabilistic,
social-based, context-aware...), that all trade off between a high
number of copies propagating in the network (hence a large network
load) and a long delay before delivery~\cite{ieeecomst15chakchouk}.
In all cases, the messages that propagate in the network are assumed
to carry immutable data as a payload, which makes these routing
protocols inadequate to handle the states of state-based CRDT
replicas.

Only a few papers have addressed the problem of sharing mutable data
in OppNets. An early proposal of shared content editing has been developed
in~\cite{chants14karkkainen}. Revision control mechanisms are used
to merge copies of a shared content when two nodes of the OppNet come
into contact, but the proposed merging mechanism still requires user
intervention to resolve conflicts. In~\cite{icta17alsulami}, an
adaptation of an Operational Transformation (OT) algorithm (OPTIC~\cite{coordination09imine})
for OppNets is evaluated. This adaptation exploits a forwarding protocol
(either Epidemic Routing~\cite{tr00vahdat} or Prophet~\cite{sapir04lindgren})
to exchange updates among groups of editing nodes. The reported evaluation
shows that whichever forwarding protocol is being used, data convergence
can be slow, with a large number of messages exchanged between peers.
Besides, message loss may occur, hence precluding convergence, due
to the cache management policy implemented in the forwarding protocols:
some messages sometimes get purged from caches before having reached
their destination.

More recently, several attempts to implement op-based CRDTs in OppNets
have been made. Costea et al. present in \cite{igi16costea} and~\cite{ccpe17costea}
approaches in which Logoot~\cite{icdcs09weiss} serves as the basis
of an op-based CRDT. A communication layer is assumed to provide causal
dissemination of the CRDT operations network-wide. These approaches
are evaluated using either Epidemic Routing~\cite{tr00vahdat} (plain
epidemics), or ONSIDE~\cite{noms14ciobanu} (socially-aware and interest-based
dissemination). Causal barriers~\cite{ccpe17costea} or CBCAST~\cite{igi16costea}
provide causal order. The experiments reported in \cite{igi16costea}
and~\cite{ccpe17costea} lead to the same observations as those made
in~\cite{icta17alsulami}: the broadcast of each operation yields
a high cost in terms of message storage and network load. Besides,
message loss can occur, and since it cannot be ignored, works in \cite{igi16costea}
and~\cite{ccpe17costea} propose to compensate this loss at application
level. 

In \cite{icoin18robin} is presented a collaborative editing system
that runs on Android and that is meant to be deployed on a delay-tolerant
network, using IBR-DTN~\cite{wowkivs11schildt} for message forwarding.
The type of CRDT considered is an op-based CRDT, also derived from
Logoot~\cite{icdcs09weiss}. Again, the evaluation of this approach
shows the efficiency problem inherent in the use of op-based CRDTs
coupled with a network-wide dissemination protocol.

As explained in Section~\ref{sec:introduction}, op-based CRDTs require
reliable (and sometimes causal) broadcast among all replicas. They
are thus mostly suited to be used in situations where good connectivity
is guaranteed between the replicas, which is hardly the case in OppNets.
Besides, deploying op-based CRDTs in OppNets requires the continuous
broadcast of many small messages (one for each update applied to a
replica), which tends to quickly saturate the caches of mobile carriers,
and ultimately lead to message loss and failure to converge. In any
case, using op-based CRDTs in an OppNet yields a high cost in terms
of network load.

Unlike op-based CRDTs, state-based CRDTs can tolerate network partitions
and message loss, and they do not require the reliable broadcast of
messages since the replicas can synchronize pairwise. State-based
CRDTs are thus ideally suited to be used in OppNets.

A small-scale real-life experiment involving the use of state-based
CRDTs in an OppNet has been reported in~\cite{ubicomm23maheo}. In
this experiment, the laptops of the members of a small research team
were configured to operate as a small OppNet, hosting replicas of
a shared document edited asynchronously by the team members. No network-wide
communication was required as replica states (or more specifically,
delta-states) were exchanged only between pairs of neighbor laptops
upon radio contact. The convergence of all replicas was ensured thanks
to the transitivity obtained via successive pairwise synchronizations,
and the storage and transmission costs observed on all participating
laptops remained acceptable.

In the above-mentioned experiment, but also in all the papers dealing
with OT or op-based CRDTs in OppNets, it is commonly assumed that
all the nodes in the network participate in the distributed application
by hosting a replica. In the present paper, the prime assumption is
that only a fraction of the network nodes host the replicas of a given
CRDT, but those that do not host replicas can yet serve as relays
to assist in the synchronization of replicas.

\section{\label{sec:System_model}System model}

\begin{figure}[h]
  \rule[0.5ex]{\columnwidth}{1pt}
  \includegraphics[width=8.5cm]{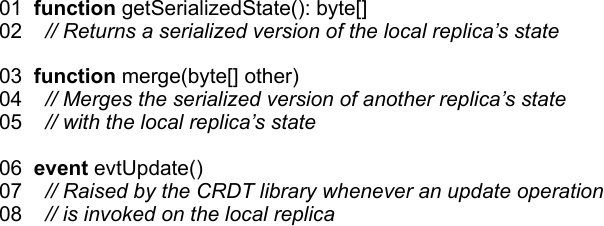}\par 
  \rule[0.5ex]{\columnwidth}{1pt}
  \caption{\label{fig:basic_functions}Definitions of the basic functions expected from any CRDT library}
\end{figure}

The relay-based synchronization system (RBSS) defined in this paper
is meant to interface seamlessly with preexisting state-based CRDT
libraries, allowing the replicas based on these libraries to synchronize
via relays, with very little or no modification of the libraries.
Whatever the CRDT library considered, and whatever the kind of CRDT
considered (register, list, map, etc.), we assume that the library's
API makes it possible to:
\begin{enumerate}
\item get a serialized version of a replica's state;
\item merge a replica's state with the serialized version of another replica's
state;
\item get notified whenever an update is issued locally.
\end{enumerate}
The serialized version of a replica's state is typically a simple
byte array, that can be stored in a file to be reloaded later, or
that can be sent to another replica to be merged there with this replica's
state. In the remainder of this paper we use the term ``serialized
version of the state'', or more simply ``serialized state'', to
distinguish this representation of the replica's state from its internal
representation, which is typically runtime-dependent and may neither
be storable and restorable as is, nor mergeable with another replica's
state. To the best of our knowledge all current implementations of
CRDT libraries provide a function that can return a replica's state
in a storable and/or transmittable form, and another function that
can instantiate a replica's state from this storable and/or transmittable
form. Besides, it is usually possible to set a listener that gets
notified whenever an update operation is issued on the local replica.
Figure~\ref{fig:basic_functions} shows the signatures of the two
basic functions, and of the event, that are needed to satisfy assumptions
1), 2) and 3). We believe that any CRDT library already provides similar
functions and events in their API, or at least provides material with
which they could be readily implemented. 

Note that the byte array returned by function{\small{} }\textsf{\small{}getSerializedState()}
(in Fig.~\ref{fig:basic_functions}) does not need to make sense
for the RBSS. Its content may be encoded in any way, depending on
the type of CRDT considered and on the library considered. It may
even be signed and/or encrypted by the CRDT library (or by a third-party
encryption module), thus ensuring that the replicas can trust the
states they exchange, even though these states may be transported
by relays rather than being obtained directly from peer replicas (this
topic is discussed further in Section~\ref{sec:Discussion}).

Likewise, the nature of the update operation signalled by event
\textsf{\small{}onEvtUpdate()} does not need to be specified. The RBSS simply
needs to be notified when an update operation is issued on the local
replica. It does not need to know any detail about this operation.

Besides interfacing with a CRDT library, the RBSS must also rely on some
communication middleware to ensure the synchronization of replicas and relays
when they get in radio contact. We assume this communication middleware takes
care of neighbor discovery on the wireless channel, and ensures a reliable
transmission of messages between neighbor nodes
(i.e. any message is delivered unaltered and once only, or not delivered).  The
radio link between two neighbor nodes may however be disrupted at any time, for
example when the nodes move away from each other, and thus cease to be
neighbors. A synchronization between two neighbor nodes may thus be interrupted
before reaching completion.

Note that we do not assume that multi-hop forwarding is supported
in any way by the communication middleware. The synchronization protocols
defined in the next section only rely on direct opportunistic interactions
between pairs of neighbor nodes.

\section{\label{sec:Synchronization_algorithms}Synchronization protocols
for CRDTs in OppNets}

The synchronization process must be different, depending on whether
the nodes that synchronize are two replicas, two relays, or a replica
and a relay. Three synchronization protocols are thus required, that
are presented in this section.

Note that in each of these protocols we always consider the synchronization
of the replicas of a single CRDT instance, for the sake of clarity.
In a real distributed system, several instances may of course coexist
in the system (for example several registers, maps, lists, etc. at
the same time), so any actual implementation of CRDT replicas and
relays should be able to differentiate these instances, and process
them accordingly.

Besides, the protocols presented below are slightly simplified for the
sake of clarity. Indeed, while presenting these protocols we
deliberately overlook the fact that a radio contact between two nodes
may be broken before a synchronization completes. We also overlook
the fact that a node may be in contact with several other nodes at the
same time, and thus have to synchronize with all these nodes
simultaneously. Such considerations are taken into account in
Section~\ref{par:enhancements}, which provides details about what
must be done to ensure that pairwise synchronizations of replicas
and relays perform satisfactorily when multiple contacts occur at the
same time and place.

\subsection{\label{subsec:rep-rep-protocol}Synchronization between two replicas}

Algorithm~\ref{alg:rep-rep-algo} presents the code required to ensure
the synchronization of two replicas. On each replica node the RBSS
maintains a version vector (line~01). Whenever an update operation
is issued locally by the CRDT library, the counter corresponding to
the self entry in the version vector is incremented to account for
this update (lines~03$-$04). When a new peer is detected, and this peer
happens to be another replica, the version vector of the local replica
is sent to the peer (line~07). In an OppNet, a peer is necessarily
a neighbor node, that has just been detected as being in the radio
range of the local node. Yet we use the term ``peer'' rather than
``neighbor'' in the algorithms defined in this paper, because these
algorithms could also be used out of the realm of OppNets (in Internet-based
networks for example, as discussed in Section~\ref{sec:Discussion}).

\begin{algorithm}[h]
\includegraphics[width=7cm]{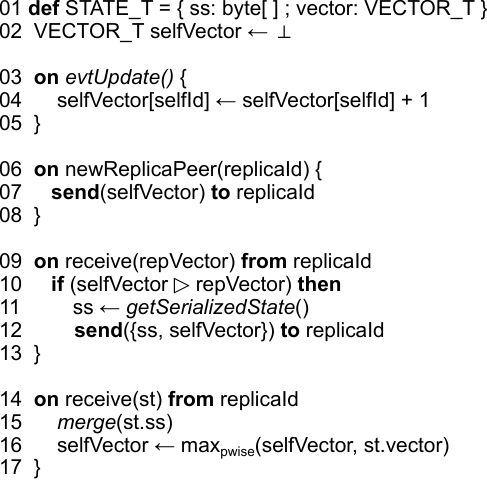}\par
\caption{\label{alg:rep-rep-algo}Replica-replica synchronization protocol}
\end{algorithm}

Upon receiving the version vector of a peer replica, the receiver
compares the received vector with its own vector in order to determine
if it is worth sending its state to the peer (line~10). If so a serialized
version of the local state is obtained from the CRDT library, and
this serialized state is sent to the peer, together with the local
version vector (lines~11$-$12). The comparison of version vectors in
line~10 aims at determining if one vector is \emph{over} ($\vartriangleright$)
the other. We define this relation as:
 $V_{A}\vartriangleright V_{B}:$~$\exists\,i|V_{A}[i]>V_{B}[i]$.
In plain terms relation $V_{A}\vartriangleright V_{B}$ means that
``the state characterized by $V_{A}$ could be used to inflate the
state characterized by $V_{B}$''. Note that the $\vartriangleright$
relation should not be confused with the \emph{ordered relation}:
$V_{A}\prec V_{B}:$~$\forall i,V_{A}[i]\leq V_{B}[i]\,and\,\exists\,j\,|~V_{A}[j]<V_{B}[j]$.
The relation $V_{A}\vartriangleright V_{B}$ holds when $V_{A}$ accounts
for updates that are not accounted for in $V_{B}$. But the relation
$V_{B}\vartriangleright V_{A}$ can hold at the same time, in which
case $V_{A}$ and $V_{B}$ are concurrent, and the states $A$ and
$B$ are conflicting states. In contrast, $V_{A}\succ V_{B}$ (opposite
of $V_{A}\prec V_{B}$) holds when all updates in $V_{B}$ are accounted
for in $V_{A}$, and $V_{A}$ accounts for at least one update that
is not accounted for in $V_{B}$. In line~10 of Alg.~\ref{alg:rep-rep-algo},
the test is used to determine if the local replica has visibility
on update operations that are not yet viewed by the peer replica,
in which case it is worth sending the local state to the peer.

Upon receiving the state of the peer replica (actually, the serialized
version of its state, together with its version vector), the receiver
merges the received serialized state with its own state, and joins
the received vector with its own vector (lines~15$-$16). Joining two
version vectors simply consists in computing the pointwise max of
both vectors. 

The synchronization algorithm sketched in Alg.~\ref{alg:rep-rep-algo}
may be implemented from scratch, using the functions and events provided
by the CRDT library (shown in italics in the algorithm). In practice,
though, any CRDT library is likely to come with its own synchronization
protocol, using either state-based synchronization (as shown in Alg.~\ref{alg:rep-rep-algo}),
or delta-state-based synchronization. If the builtin synchronization
protocol of a CRDT library must be used while allowing relay-based
synchronization, the developer should only ensure that a version vector
is maintained on each replica, and that the version vectors are exchanged
and joined whenever two replicas synchronize (as shown in lines~04
and~07). This is required if the replicas are expected to synchronize
also with relays, as discussed below.

\begin{figure}[h]
\begin{centering}
\includegraphics[width=10.7cm]{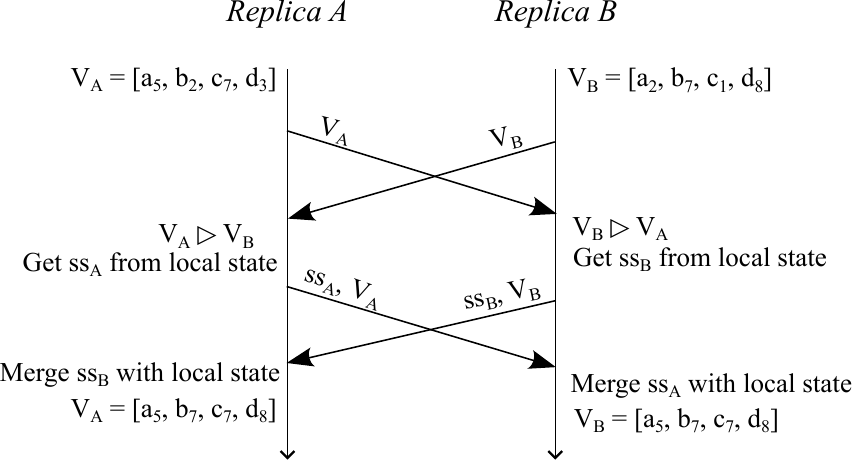}
\par\end{centering}
\caption{\label{fig:Example-rep-rep}Example of synchronization between two
replicas}
\end{figure}

Fig.~\ref{fig:Example-rep-rep} illustrates a synchronization
between two replicas A and B. At the beginning of this synchronization,
the version vector of A is $V_{A}=\left[a_{5},b_{2},c_{7},d_{3}\right]$,
and that of B is $V_{B}=\left[a_{2},b_{7},c_{1},d_{8}\right]$. Note
that in this example we present each version vector as a collection
of $k_{v}$ pairs, where $k$ is the identifier of a replica, and
$v$ is the value of the update counter known for this replica. $V_{A}$
therefore indicates that A's state accounts for ---or ``views''---
5 updates that have been issued (locally) on A itself, 2 updates issued
on B, 7 updates issued on C, and 3 updates issued on D. 

As explained above, each replica first sends its local version vector
to the peer replica, so that each receiver can determine if it is
worth sending its state to the peer. In that case, since the two states
are in conflict (i.e., $V_{A}\triangleright V_{B}$ and $V_{B}\triangleright V_{A}$),
both replicas end up sending their state to the peer, and after merging
the received state with its own, both replicas end up with the same
state, and with the same version vector.

The code shown in Alg.~\ref{alg:rep-rep-algo} is meant to minimize
the number of states exchanged when two replicas synchronize. Depending
on the relative initial states of the replicas that engage in a synchronization,
the number of serialized states transferred during that synchronization
may be zero (i.e., the replicas are already in sync), one (only one
replica can gain from the synchronization), or two (both replicas
can gain from the synchronization).

\subsection{\label{subsec:rep-rel-protocol}Synchronization between a replica
and a relay}

Algorithms~\ref{alg:rep-rel-algo-replica-side} and~\ref{alg:rep-rel-algo-relay-side}
present the code required on both sides to ensure the synchronization
between a replica and a relay. Note that unlike a replica, a relay
is unable to merge states, so it must implement a store in which multiple
conflicting states can be maintained simultaneously (line~02 in Alg.~\ref{alg:rep-rel-algo-relay-side}).
This store is simply a set of tuples, each tuple combining a version
vector and the associated serialized version of the issuer replica's
state. Besides maintaining a store of conflicting states, the relay
also maintains an aggregate vector (\emph{vagg}), whose role is explained
below.

\begin{algorithm}[h]
  \includegraphics[width=7cm]{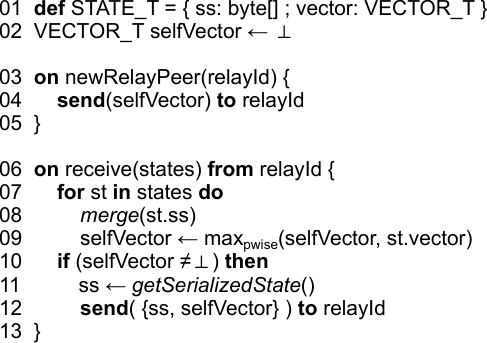}\par 
  \caption{\label{alg:rep-rel-algo-replica-side}Replica-relay synchronization protocol (replica side)}
\end{algorithm}

\begin{algorithm}[h]
  \includegraphics[width=9.6cm]{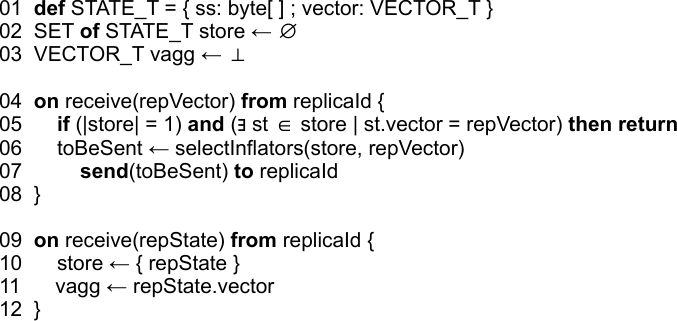}\par 
  \caption{\label{alg:rep-rel-algo-relay-side}Replica-relay synchronization protocol (relay side)}
\end{algorithm}

Whenever a new peer is detected by a replica, and this peer happens to
be a relay, the replica sends its current version vector to the relay
(line~04 in Alg.~\ref{alg:rep-rel-algo-replica-side}). Upon receiving
this vector, the relay selects in its store the states it should send
to the replica (lines~06$-$07 in Alg.~\ref{alg:rep-rel-algo-relay-side}).
This selection is achieved by calling function \textsf{\small
  selectInflators()}, which is presented in
Alg.~\ref{alg:select-dominating} and explained in
Section~\ref{subsec:selectInflators}. Basically, this function is meant
to return a minimal subset of the states contained in the relay's
store that can be used to inflate the replica's state. Note that this
subset may be empty, and the relay may then send an empty set of
states to the replica, so the replica can reply by sending its own
state back, as explained below. The only case where the relay should
send nothing to the replica is when the relay's store contains only
one state, and this state is equal to the state of the replica, in
which case both hosts are already synchronized (line~05 in
Alg.~\ref{alg:rep-rel-algo-relay-side}).

Once function \textsf{\small selectInflators()} has returned on the
relay, the selected states and their version vectors can be sent to
the replica (line~07 in Alg.~\ref{alg:rep-rel-algo-relay-side}).

\begin{figure}[h]
\begin{centering}
\includegraphics[width=10.7cm]{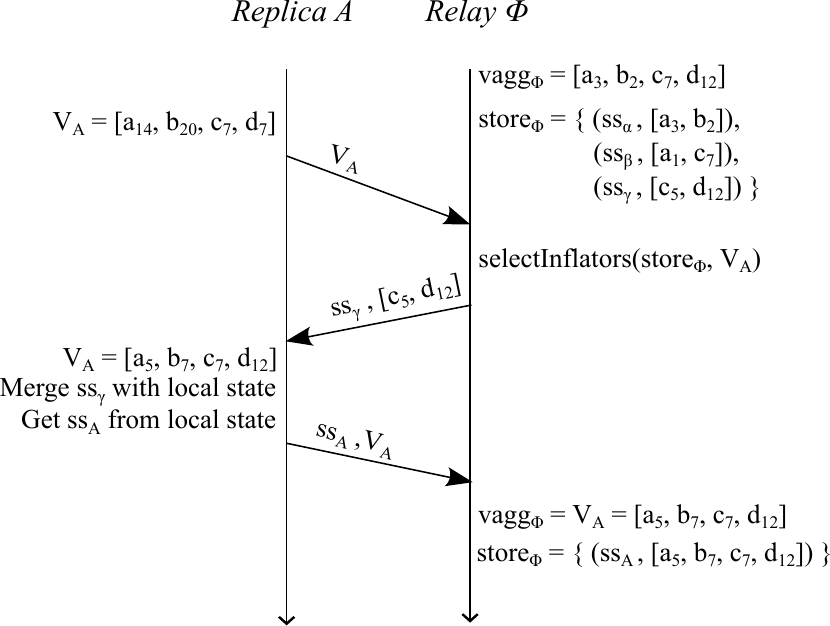}
\par\end{centering}
\caption{\label{fig:Example-rep-rel}Example of synchronization between a replica and a relay}
\end{figure}

Fig.~\ref{fig:Example-rep-rel} illustrates a synchronization between replica A
and relay \textgreek{F}. In this example the version vector of A is initially
$V_{A}=\left[a_{5},b_{2},c_{7},d_{7}\right]$. Relay \textgreek{F} holds several
conflicting states in its store, each with its own version vector, respectively
$\left[a_{3},b_{2}\right]$, $\left[a_{1},c_{7}\right]$ and
$\left[c_{5},d_{12}\right]$. As explained above the synchronization between a
replica and a relay begins with the replica sending its own version vector to
the relay. The relay then determines which of the states it holds should be
sent to the replica. In the example considered here,
$\left[a_{3},b_{2}\right]\ntriangleright V_{A}$, and
$\left[a_{1},c_{7}\right]\ntriangleright V_{A},$ so sending the corresponding
states to A would not help it to inflate its own state.  In contrast,
$\left[c_{5},d_{12}\right]\vartriangleright V_{A}$ (because of the value of
item $d$) so the corresponding set constitutes the minimal subset of states to
send. This set allows A to receive updates it has never received before, namely
all the updates issued by D from $d_{8}$ to $d_{12}$.
 
Upon receiving a state from the relay, the replica merges this state
with its own local state, and joins the received version vector with
its own (lines~07$-$09 in Alg.~\ref{alg:rep-rel-algo-replica-side}).
Once the replica has received all contributions from the relay and
updated its local state accordingly, the replica sends its new state
to the relay (lines~11$-$12 in Alg.~\ref{alg:rep-rel-algo-replica-side}).

Note that this transmission occurs only if the replica's state is not
empty (line~09). Thus, as long as no update is visible yet to the
replica (the replica's state is empty, and nothing new has been
received from the relay), nothing is sent to the relay.

Upon receiving the new state of the replica, the relay places this state in the
store, where it replaces all the states that may have been in the store
previously (line~10 in Alg.~\ref{alg:rep-rel-algo-relay-side}).  It also uses
the received version vector to reset its own aggregate vector (line~11). In our
example the relay ends up with a store that contains only the new state
received from the replica:
$store_{\Phi}=\{(ss_{A},\thinspace\left[a_{5},b_{7},c_{7},d_{12}\right])\}$.

As a general rule, a synchronization between a replica and a relay
implies that either one or two version vectors are exchanged between
both hosts. Besides, the replica first receives a number of states
from the relay (but not necessarily the whole content of the relay's
store, and sometimes no state at all), merges these states with its
own single state, and returns the result to the relay. At the end
of this synchronization process, the relay always ends up with only
one state in its store: the same as that of the replica.

\subsection{\label{subsec:rel-rel-protocol}Synchronization between two relays}

Algorithm~\ref{alg:rel-rel-algo} shows the code required to ensure
the synchronization of two relays. Whenever a new peer is detected
by a relay, and this peer happens to be another relay, each relay
sends its current aggregate version vector to its peer (lines~04$-$05 in
Alg.~~\ref{alg:rel-rel-algo}). The aggregate version vector characterizes
the whole set of states that are contained in the relay's store. In
essence, it is equivalent to the version vector of the single state
that would be maintained by the relay if the relay was able to merge
all the states contained in its store (which it is not). Computing
the aggregate version vector of a relay simply comes down to computing
the pointwise max of all the version vectors associated with the states
maintained in its store. 

\begin{algorithm}[h]
  \includegraphics[width=7cm]{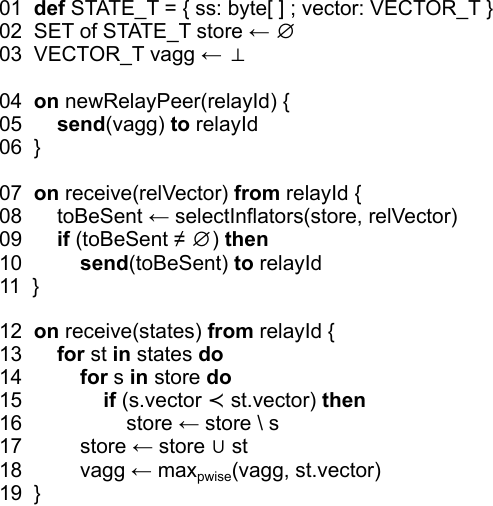}\par 
  \caption{\label{alg:rel-rel-algo}Relay-relay synchronization protocol}
\end{algorithm}

\begin{figure}[h]
\begin{centering}
\includegraphics[width=13.8cm]{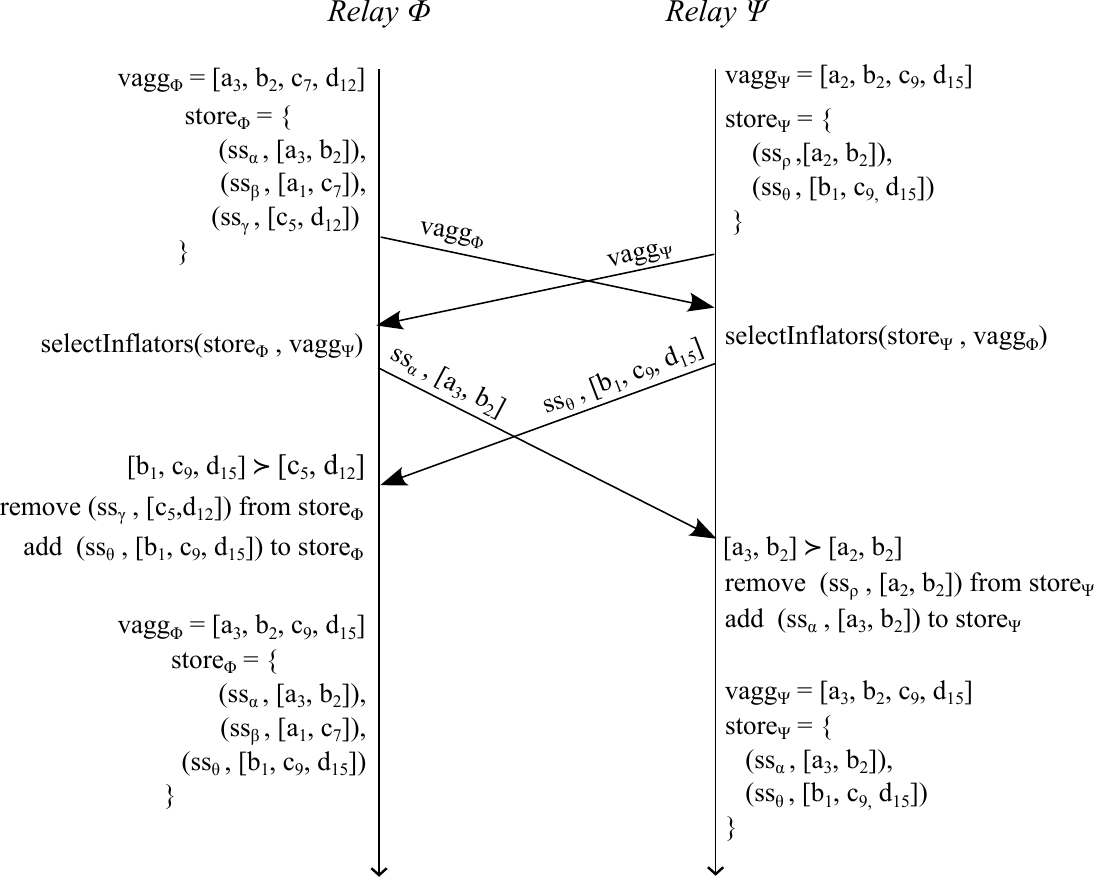}
\par\end{centering}
\caption{\label{fig:Example-rel-rel}Example of synchronization between two
relays}
\end{figure}

Fig.~\ref{fig:Example-rel-rel} illustrates a synchronization 
between two relays \textgreek{F} and $\Psi$. In this example \textgreek{F}
initially holds three conflicting states, whose version vectors are
$\left[a_{\boldsymbol{3}},b_{\boldsymbol{2}}\right]$, $\left[a_{\boldsymbol{1}},c_{\boldsymbol{7}}\right]$
and $\left[c_{5},d_{12}\right]$. Likewise, ${\color{blue}{\normalcolor \Psi}}$
initially holds two conflicting states, whose version vectors are
$\left[a_{2},b_{2}\right]$ and $\left[{\color{blue}\boldsymbol{{\normalcolor {\normalcolor b_{1},c}}}}_{9},d_{15}\right]$.
The aggregate version vectors of both relays are respectively $vagg_{\phi}=[a_{3},b_{2},c_{7},d_{12}]$
and $vagg_{\psi}=[a_{2},b_{2},c_{9},d_{15}]$. These vectors are exchanged
between the relays at the beginning of the synchronization procedure.

Upon receiving the aggregate version vector from its peer, each
receiver uses function \textsf{\small selectInflators()} to select
among the states contained in its store which of these states it must
send to the peer (lines~08$-$10 in Alg.~\ref{alg:rel-rel-algo}). Note
that there is no need to send anything if the set of selected states
is empty, hence the test in line~09.

In our example, upon receiving $vagg_{\Phi}$, relay $\Psi$ observes that
$\left[\boldsymbol{a}_{\boldsymbol{2}},\boldsymbol{b}_{2}\right]\ntriangleright
V_{agg}(\Phi)$, but
$\left[\boldsymbol{b_{1}}\boldsymbol{,c}_{9},d_{\boldsymbol{15}}\right]\vartriangleright
vagg_{\Phi}$ (because of the values of items c and $d$). So the state
corresponding to
$\left[\boldsymbol{b_{1}}\boldsymbol{,c}_{9},d_{\boldsymbol{15}}\right]$
constitutes the minimal subset of states to be sent to $\Phi$.

Likewise, upon receiving $vagg_{\psi}$, relay $\Phi$ observes that
$\left[\boldsymbol{a}_{\boldsymbol{3}},\boldsymbol{b}_{2}\right]\vartriangleright V_{\psi}$,
whereas $\left[\boldsymbol{a}_{\boldsymbol{1}},c_{7}\right]\ntriangleright vagg_{\psi}$,
and $\left[c_{5},d_{12}\right]\ntriangleright vagg_{\psi}$, so only
the first state must be sent to $\Psi$.

Upon receiving the states sent by the peer relay, the receiver can add
these states to its store (line~17 in~Alg.\ref{alg:rel-rel-algo}), while
purging the store from earlier states that are now rendered obsolete
by the new states (lines~14$-$16). Note that the test in line~15 relies
on the ordered relation ($\prec$).

In our example, upon receiving the state whose version vector is
$\left[\boldsymbol{b_{1}}\boldsymbol{,c}_{9},d_{\boldsymbol{15}}\right]$, relay
$\Phi$ can remove $\left[c_{5},d_{12}\right]$ from its store, since
$\left[\boldsymbol{b_{1}}\boldsymbol{,c}_{9},d_{\boldsymbol{15}}\right]\succ\left[c_{5},d_{12}\right]$.
Likewise, upon receiving $\left[\boldsymbol{a_{3}}\boldsymbol{,b}_{2}\right]$,
relay $\psi$ can remove $\left[a_{2},b_{2}\right]$ from its store, since
$\left[\boldsymbol{a_{3}}\boldsymbol{,b}_{2}\right]\succ\left[a_{2},b{}_{2}\right]$.

Finally, the aggregate version vector of a relay is also updated whenever
a new state is placed in its store (line~18), so this vector is kept
up-to-date and will be ready when the relay engages in a new synchronization.

At the end of the synchronization between relays $\Phi$ and $\Psi$,
$store_{\Phi}=\{$$(ss_{\alpha},\thinspace\left[a_{\boldsymbol{3}},b_{\boldsymbol{2}}\right]),(ss_{\beta},\thinspace\left[a_{1},\boldsymbol{c_{7}}\right]),$
$(ss_{\theta},\thinspace\left[b_{1},c_{9},d_{15}\right])\}$ and $store_{\Psi}=\left\{ (ss_{\alpha},\left[a_{3},b_{2}\right]),\,(ss_{\theta,}\left[\boldsymbol{b_{1},c}_{9},d_{15}\right])\right\} $.
Although relays $\Phi$ and $\Psi$ have just synchronized, their
stores still contain different states. Their aggregate version vectors
are similar, though: $vagg_{\Phi}=vagg_{\psi}=\left[a_{3},b_{2},c_{9},d_{15}\right]$,
which confirms that both stores account for the same sets of updates.
A replica, after synchronizing with either $\Phi$ or $\Psi$, would
end up with the same state.

As a general rule, a synchronization between two relays implies that
several states may be transmitted, as illustrated in the former example.
The number of states actually transmitted in each direction is kept
at a minimum, though, thanks to the selection performed in line~08
(Alg.~\ref{alg:rel-rel-algo}): only states that are required
to inflate the peer's store are transmitted.

The purge performed in lines~15$-$16 (in Alg.~\ref{alg:rel-rel-algo})
is meant to ensure that the number of states maintained in a relay's
store is also kept at a minimum. Besides, the following assertions
are verified at any time:
\begin{enumerate}
\item a store can only contain conflicting states (i.e., states that cannot
be causally ordered by the relay);
\item the number of such states is bounded by the number of replicas in
the system.
\end{enumerate}

\paragraph*{Proof of assertion 1}
Assume the store $S$ of a relay actually only contains conflicting
states, i.e., $\forall s,t\in S,s\neq t:V_{s}\parallel V_{t}\,\Leftrightarrow\,(V_{s}\vartriangleright V_{t}\,\wedge\,V_{t}\vartriangleright V_{s}).$
In plain terms: for every pair of states contained in the store, each
state accounts for at least one update that has not yet been accounted
for in the other state. This statement is of course verified after
a relay has synchronized with a replica, for in that case the store
only contains one state (see line~10 in Alg.~\ref{alg:rep-rel-algo-relay-side}).
Assume a relay receives a new state $u$ from a peer relay and must
add this state to its store. State $u$ has been selected by the peer
because it accounts for updates that are not yet accounted for in
the local store (line~47 in Alg.~\ref{alg:select-dominating}). Thus $\forall t\in S:V_{u}\vartriangleright V_{t}$.
By purging the store from any state that is causally dominated by
$u$ (lines~13$-$14 in Alg.~\ref{alg:rel-rel-algo}), the local relay removes $\forall t\in S:V_{u}\succ V_{t}$.,
i.e., $\forall i,V_{u}[i]\geq V_{t}[i]\,\wedge\,V_{u}\vartriangleright V_{t}.$
Any state $s$ that is \emph{not} purged from the local store is thus
a state such that $\forall i,V_{u}[i]\ngeq V_{s}[i]\,\Leftrightarrow\,\exists j|V_{u}[j]<V_{s}[j]\,\Leftrightarrow\,V_{u}\vartriangleleft V_{s}$.

So after adding $u$ to the store we have~ $\forall s\in S,s\neq u:$
 $(V_{s}\vartriangleright V_{u}\,\wedge\,V_{u}\vartriangleright V_{s})\,\Leftrightarrow\,V_{s}\parallel V_{u}$.
Any state added to the store is thus a state that is in conflict with
every other state maintained in the store: at any time the store only
contains conflicting states, that is, states that cannot be causally
ordered by the relay, since the relay is unable to merge conflicting
states.

\paragraph*{Proof of assertion 2}
Since all the states issued by a replica are causally ordered, and
since the relay cannot hold causally ordered states, the store can
contain at most one state issued by each replica. The purge performed
in lines~15$-$16 (in Alg.~\ref{alg:rel-rel-algo}) therefore ensures
that the number of states maintained in a relay's store is kept at
a minimum, and this number can never exceed the number of replicas
in the system.

\subsection{\label{subsec:selectInflators}Selection of states to inflate a peer}

Whenever a relay synchonizes with a replica or with another relay, it must
select which of the states contained in its own store could contribute
to inflate that peer's state. This selection is achieved by calling function \textsf{\small selectInflators()}, whose code is detailed in
Alg.~\ref{alg:select-dominating}.

In this function, all the states whose version vector is \emph{over}
that of the peer are initially considered as candidates to be sent
to the peer (line~47 in Alg.~\ref{alg:select-dominating}). The
version vector \textsf{\small vvcand} corresponding to the aggregation
of these candidates' vectors is computed (line~48), and based on
\textsf{\small vvcand} and the peer's version vector \textsf{\small
  vv}, a target version vector \textsf{\small vvtarget} is determined
(line~49). In this target vector, only the values of the counters that can be
inflated by sending at least one state to the peer are
maintained. The other counters are all set to 0. The remainder of the
selection process comes down to selecting a subset of the candidates
that should be sent to the peer so its own version vector can reach
all the non-zero values registered in \textsf{\small vvtarget}. This
is a typical instance of the set cover problem, with a universe that
consists of the non-zero counters registered in \textsf{\small
  vvtarget}, each candidate state allowing to cover all or part of
this universe.

Since the set cover problem is NP-complete, several approximate
solutions have been proposed in the literature. In the following we
use a slightly refined version of the greedy algorithm, which is known
to provide polynomial time approximation of set covering, while having
a reasonably lightweight footprint~\cite{appalgo97hochbaum}. The pure
greedy approach consists in always selecting, among the remaining
candidates, a candidate that covers the largest collection of as-yet
uncovered elements of the universe (i.e., non-zero counters in
\textsf{\small vvtarget}).  The refinement we introduce consists in
identifying first (line~50) the states that must absolutely be
selected to cover elements of the universe. This is performed by
calling \textsf{\small getSingleInflators()}, which selects states
that can contribute by inflating a counter that would not be inflated
by any other state. The selection of these ``single inflators'' is
performed by considering each non-null counter value in the target
version vector, and selecting a state only if it can uniquely reach
this value (lines~23$-$26 in Alg.~\ref{alg:select-dominating}). Note that
whenever such a state is selected, the target version vector is
updated by ``masking out'' (i.e., setting to 0) all counter values
that can be reached thanks to the selected state (line~27).

Once the ``single inflators'' have all been selected, they are not
candidates anymore (line~51). Unless the target version vector is now
empty, the second step of the selection process is initiated, based on
the greedy algorithm mentioned above. This step consists in looking
iteratively for each candidate that can bring the maximal benefit to
the peer (line~53). This is a state that ``covers'' the highest number
of non-zero values of the target version vector (lines~39$-$44). Again,
when such a state is selected, it is removed from the set of remaining
candidates (line~55), and the target version vector is ``masked out''
accordingly (line~56). The selection process is complete when the target version
vector is empty, i.e., a subset of candidates has been determined that
can contribute to fully inflate the peer's state.

\begin{figure}[h]
\begin{centering}
\includegraphics[width=12cm]{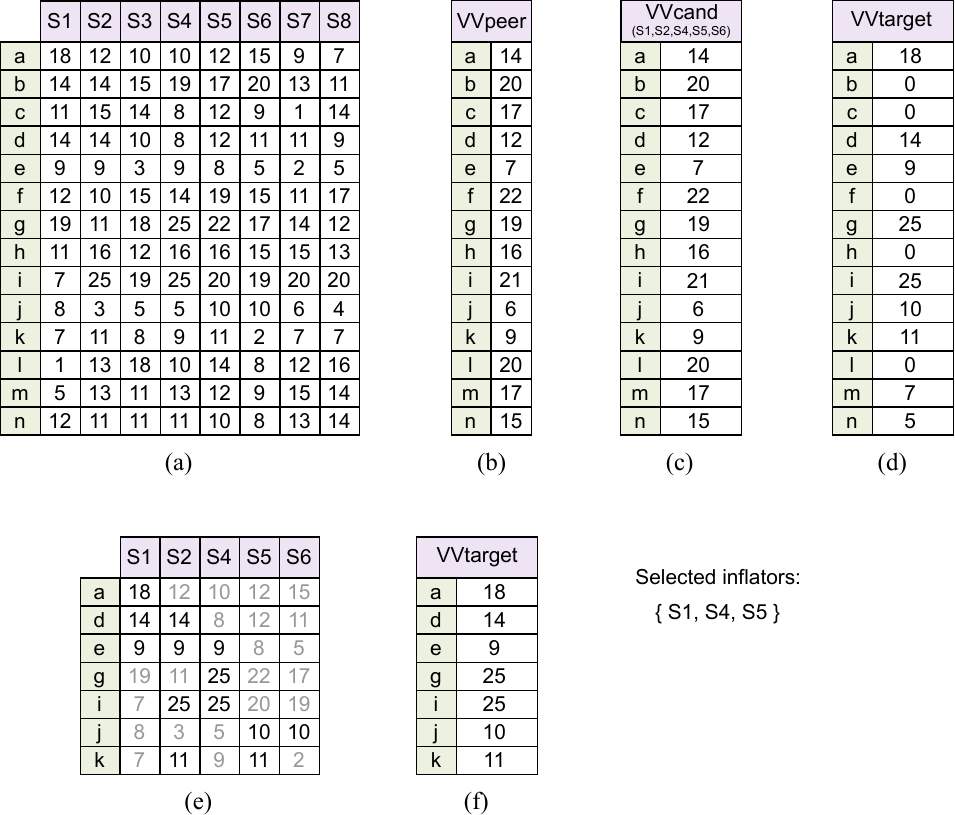}
\par\end{centering}
\caption{\label{fig:Ilustration-selection}Illustration of the selection of states to be transferred by a relay to a peer}
\end{figure}

Fig.~\ref{fig:Ilustration-selection} provides an illustration of the
selection process described above. In this figure we consider a
scenario that involves 14 replicas (\emph{a} to \emph{n}), and a
relay whose store contains 8 conflicting states (\emph{S1} to
\emph{S8}). Fig.~\ref{fig:Ilustration-selection}{\textcolor{blue}a} shows the version
vectors of these states.

At some time the relay gets into contact with a peer, and receives
this peer's version vector \emph{VVpeer}. As explained above, the set
of candidate states is computed first (line~47 in
Alg.~\ref{alg:select-dominating}). In this example, the set of
candidates capable of inflating the peer's state is
$\{S1,S2,S4,S5,S6\}$.

The aggregate version vector for these candidates \emph{VVcand} is
computed (line~48), after which the target version vector
\emph{VVtarget} can be computed as well (line~49). In \emph{VVtarget},
the counters whose value is non-zero define the universe to be covered
while selecting which candidate states should be sent to the peer.

Fig.~\ref{fig:Ilustration-selection}{\textcolor{blue}e}
and~\ref{fig:Ilustration-selection}{\textcolor{blue}f} show the
parameters of the problem to be solved when considering only the
candidate states (with non-maximal values greyed out) and
\emph{VVtarget}'s non-zero counters.

\begin{algorithm}[H]
  \includegraphics[width=10cm]{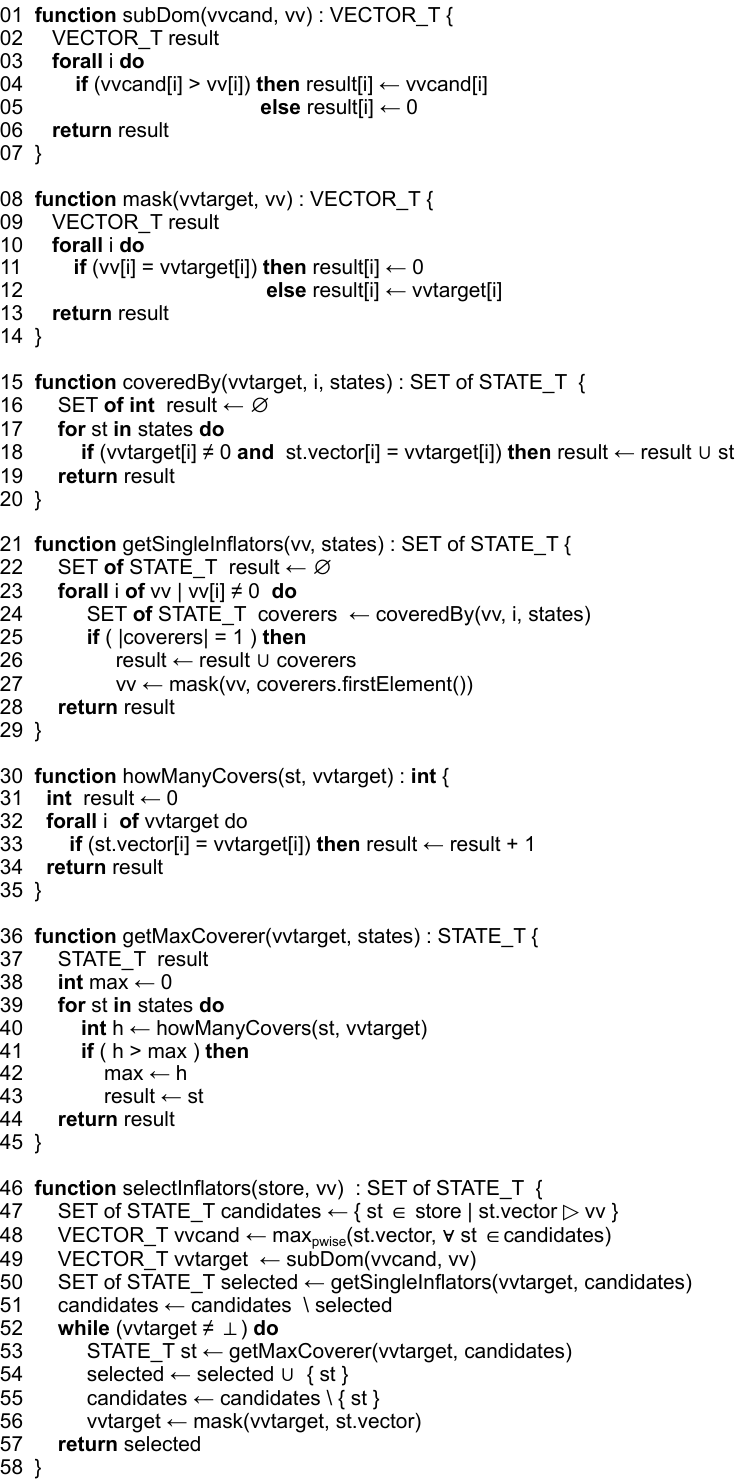}\par 
  \caption{\label{alg:select-dominating}Selection of the minimum subset of inflators (relay side)}
\end{algorithm}

Once \emph{VVtarget} has been determined, the selection of ``single
inflators can be performed (line~50). In the example considered in
Fig.~\ref{fig:Ilustration-selection}, S1 must be selected since it is
the only candidate that covers target $a_{18}$. Note that once S1 is
selected, targets $d_{14}$ and $e_{9}$ are covered as well. S4 must
likewise be selected as it is the only candidate that covers $g_{25}$
(together with $i_{25}$). As the selection of ``single inflators''
completes, S1 and S4 have been selected, targets
$\{a_{18},d_{14},e_{9},g_{25},i_{25}\}$ are covered, and only targets
$\{j_{10},k_{11}\}$ are not covered yet.  The greedy algorithm is then
used (lines~52$-$56) to select iteratively the next candidates capable
of covering the remaining target counters. In our example it turns out
that $\{j_{10},k_{11}\}$ can be covered by S5, so the final covering
subset is $\{S1,S4,S5\}$.

In this selection process, the advantage of selecting single inflators
first is that it reduces the population of candidates that must be
considered by the greedy algorithm. Besides, it can sometimes lead to
better approximations of the ideal set covering. For example, using
only the greedy algorithm with the example depicted in
Fig.~\ref{fig:Ilustration-selection} would lead to a 4-state solution
(either $\{S2,S1,S4,S5\}$ or $\{S2,S1,S4,S6\}$), instead of the
3-state solution $\{S1,S4,S5\}$ obtained by selecting single inflators
first.

The computational complexity of the selection process is
O(\emph{n.m}), where \emph{n} is the number of replicas in the
scenario considered (which determines the size of version vectors),
and \emph{m} is the number of states contained in a relay's store
(which determines the number of candidates). Although \emph{m} is
bounded by \emph{n}, the experiments presented in
Section~\ref{sec:Experimentation} show that \emph{m} is often far
smaller than \emph{n}, sometimes by several orders of magnitude. The
computational time required for a relay to determine which states it
must send to a neighbor therefore remains reasonable, even when
\emph{n} is large.

\subsection{Dealing with contact loss and multiple neighbors\label{par:enhancements}}

As explained earlier, the synchronization algorithms presented above describe
how a replica or a relay should behave in order to synchronize with a single
peer. They do not take into consideration the fact that a radio contact between
two peers may be interrupted at any time, or the fact that a replica or a relay
may have to interact with several peers concurrently. Such situations are
considered below, and several modifications are proposed to the synchronisation
algorithms. The resulting enhanced algorithms are presented
in Appendix~\ref{sec:enhanced_algorithms}.

\paragraph{Loss of radio contact during a synchronization}

In an OppNet, a radio contact between two peers can be lost at any
time, typically when these peers move out of radio range. A
synchronization can thus be interrupted before its completion. The
algorithms presented above tolerate such interruptions: even if both
peers do not reach the same state because of contact loss, each peer
will still have had a chance to inflate its local state (at least
partially) based on contributions received from the other
peer. However, in Alg.~\ref{alg:rep-rel-algo-relay-side} we have
presented a mechanism whereby a relay sends to a replica, in a single
message, all the states it has selected for that replica (line~07 in
Alg.~\ref{alg:rep-rel-algo-relay-side}). If the radio contact between
both peers is lost while the relay is sending this single message,
nothing is received by the replica.  A better approach consists in
sending the selected states one by one, thus allowing that some of
these contributions at least can reach the replica (lines~16$-$21 in
Alg.~\ref{alg:full-relay-algo}). Each state sent by the relay to a
replica must then be accompanied by a flag (line~21 in
Alg.~\ref{alg:full-relay-algo}) that allows the replica to determine if
it has received the last contribution from the relay, in which case it
is time for the replica to return its new state to the relay
(lines~31$-$34 in Alg.~\ref{alg:full-replica-algo}). With this approach,
if the radio contact is lost before the synchronization completes, the
relay will never receive anything from the replica (so its store will
remain unchanged), but the replica will have managed to inflate its
own state partially, based on the contributions received from the
relay. Note that this strategy is based on the assumption that each
state is quite large, and the transmission of a state quite costly, so
the idea is that the replica should only send its state to the relay
\emph{after} it has received and incorporated all contributions from
the relay. If the states are not too large, an alternative strategy
would be for the replica to return its new state after receiving each
contribution from the relay, so the relay's store would be updated
incrementally during the synchronization.

The idea of having a relay send the states it has selected for the
peer in several distinct messages rather than in a single large one also
applies when the peer is another relay. The single transmission in
Alg.~\ref{alg:rel-rel-algo} (line~10) can thus be replaced by an
iteration to transfer the selected states one by one, so the receiver
can incorporate each contribution as and when received.

\paragraph{Concurrent synchronizations}
In Alg.~\ref{alg:rep-rep-algo} to \ref{alg:rel-rel-algo} we have
implicitly assumed that two peers (relays or replicas) can complete
their synchronization independently, in a somewhat atomic manner,
without any disruption by other replicas or relays. In practice, it
may happen that a peer get involved in several synchronizations (with
as many peers), and has to proceed with these synchronizations
simultaneously.

In such a situation it is important to prevent a regression of a
relay's store. This may for example occur when a relay \textit{A}
synchronizes simultaneously with a replica \textit{B} and another peer
(either replica or relay) \textit{C}. The relay \textit{A} may update
its store with one or several states obtained from \textit{C}, then
receive from \textit{B} a state that happens to be ``older'' than what
has just been put in the store. In
Alg.~\ref{alg:rep-rel-algo-relay-side}, a state received from a
replica is meant to replace what is contained in the store (lines~10
and~11). There is thus a risk of regression of the store. The solution
is to deal with the replica's state as if this state had been received
from a relay. Depending on whether its version vector is dominated,
dominates, or is concurrent with the local aggegate vector, the state
received from the replica can be discarded, used to replace the whole
content of the store, or inserted in the store (lines 52$-$55 in
Alg.~\ref{alg:full-relay-algo}).

Due to the interleaving of transmissions on the wireless channel, a relay may
also engage in the transmission of a collection of states to a peer (those
selected via a call to \textsf{\small selectInflators()}), and while this
sequence of transmissions is in progress the relay may receive better (i.e.,
more advanced) states from another peer and update its store accordingly. In
order to deal with such a situation, a solution is to make it possible for a
relay to maintain a set of the states it plans to send to each peer (line~05
in Alg.~\ref{alg:full-relay-algo}). This set can then be purged and refilled
whenever the content of the store is modified (line~11).

\paragraph{Transitive synchronization in stable neighborhoods}
Although our synchronization protocols are meant to operate in an
OppNet, where a node's neighborhood tends to change frequently, there
may however be periods when a few nodes stay in contact for a long
time. In Alg.~\ref{alg:rep-rep-algo} to~\ref{alg:rel-rel-algo}, the
synchronization between two peers is only triggered upon neighbor
discovery. As a consequence, an update event on a replica does not
trigger a synchronization with any peer it might already be in contact
with. Likewise, if peers \textit{A} and \textit{B} are in contact and are already
synchronized, and if \textit{A} gets a new neighbor \textit{C} and synchronizes with \textit{C},
there is no mechanism whereby \textit{A} (whose state has changed) may trigger
a new synchronization with \textit{B}.

It is however desirable to ensure that inflations to a replica's state
or to a relay's store be propagated rapidly to neighbor nodes. When
the state of a node (either replica or relay) is inflated, this node
must notify its neighbors, so they can return their version vectors
(lines~4$-$6 in Alg.~\ref{alg:full-replica-algo} and 6$-$8 in
Alg.~\ref{alg:full-relay-algo}). Based on these vectors, the
node can determine what needs to be sent to each neighbor. In essence
this approach comes down to triggering a new synchronization with each
neighbor, whenever the local state changes.

Note that whenever the state of a node is inflated, this inflation
first propagates to the current neighbors of this node, which can in
turn propagate the inflation to their own neighbors (if any),
etc.

The approach whereby each inflation to a node's state is immediately
propagated to its neighbors may yield a lot of traffic on the wireless
channel. In a situation where some latency can be tolerated, an
alternative approach consists in having each node trigger
synchronizations only periodically with its neighbors, rather than
after each inflation of its own state. Depending on the application
scenario considered, and depending on the dynamics of the OppNet
considered, a tradeoff can thus be found between the reactivity to
inflations and the communication overhead.

\section{\label{sec:Experimentation}Experimentation}

\subsection{A simulator for relay-based synchronization}

In order to observe how relays can influence the synchronization of
replicas in an OppNet, we designed a simulator that can play two scenarios
simultaneously: a contact scenario, and an application scenario. The
contact scenario defines which nodes are involved in the network considered,
what is the role assumed by each node (either replica, relay, or no
role at all), and what is the timeline of contacts between these nodes.
The application scenario defines what is the timeline of updates applied
to the replicas. Without loss of generality we assume that a single
CRDT is considered in an application scenario, so all replicas are
replicas of the same CRDT. In a fully-operational implementation of
a relay-based synchronization system it would of course be necessary
to process several CRDTs concurrently. 

The continuously changing topology of an opportunistic network is
ideally modelled as a time-varying graph. Our simulator has therefore
been developed using the GraphStream library\footnote{\url{https://graphstream-project.org/}},
which makes it possible to process time-varying graphs easily. 

\subsection{Metrics considered\label{subsec:Metrics-considered}}

In the literature dealing with CRDTs, the time it takes for replicas
to converge is usually overlooked. Indeed, state convergence in replicas
is often considered to be a never-ending process, i.e., new update
operations are issued continuously on the replicas so, in essence,
the replicas can never truly reach any final state. As explained in
Section~\ref{sec:introduction}, the SEC (Strong Eventual Consistency)
model guarantees that replicas keep converging as long as new updates
are issued, and if no new update is issued for long enough the replicas
eventually reach equivalent states. When updates are issued continuously,
as is often the case, possible metrics could be the time it takes
for updates to be viewed by all replicas, or the time it takes for
a replica to synchronize with another replica once they have decided
to synchronize. Yet even such metrics are also often overlooked in
the literature, for in most papers the replicas are assumed to synchronize
via Internet routes, which connect either clients and servers, or
peers in an overlay network. When a replica must synchronize with
another replica, it is therefore commonly assumed that the synchronization
either succeeds almost immediately, or fails altogether (because the
network may be temporarily partitioned, with the two replicas belonging
to separate parts of the network).

In an opportunistic network, a synchronization can only occur when
two nodes get in radio contact, which happens only every now and then.
Besides, the speed at which information can flow in the network is
mostly determined by the speed at which mobile nodes move, which is
of course significantly slower (by several orders of magnitude) than
the speed of radio waves. Every radio contact between two nodes is
therefore an opportunity for information (a replica's state in our
case) to ``jump'' almost instantaneously from one node to the next,
but after that jump this information resumes moving at the same speed
as the nodes that are carrying it. 

Network-wide replica synchronization in an OppNet can therefore only
be a rather slow process, at least when compared to seemingly instantaneous
Internet-based synchronization, and the main motivation for using
relays is to make it faster. How much faster, and at what cost, are
the two main questions that should be considered.

In order to characterize how much faster replicas can converge when
relays are used to assist in the synchronization, we must define metrics
that show how a replica's state deviates from an ideal reference state.
To do so, we define this reference state as being the global state,
that is, the single state that would be observed if all update operations
were issued on this state, with no significant delay. In other words,
the global state is that of a CRDT that would be centralized (instead
of being distributed in many replicas), and on which all updates would
be issued instantaneously according to the application scenario considered.
Alternatively, the global state can also be perceived as the state
of all the replicas if they were able to synchronize instantaneously,
all together, whenever an update operation is issued on any of them.

Determining how the global state progresses over time is easy, based
on the application scenario. The global state can be characterized
by a version vector of $(id,counter)$ pairs, the counter being incremented
instantaneously whenever an update operation is issued on replica
$id$. Using this version vector as a reference, it is possible to
observe while running simulations how ``late'' each replica node
is at any time with respect to the (ideal) global state. This ``lateness''
can be expressed either in time units or in terms of missing updates.

In the remainder of this paper we will call \emph{convergence latency}
(or \emph{latency} for short) the time required for a replica's state
to catch up with the global state, as it was at a given time. We will
likewise call \emph{convergence distance} (or \emph{distance} for
short) the number of updates a replica has not seen yet, although
these updates are seen from the global state.

More specifically, whenever an update operation is issued on a replica,
the version vector of the global state is updated immediately to account
for this update. The version vector of the local replica can likewise
be updated immediately, but for other replicas it will take time until
their version vector can be updated to reflect the update operation
that has just been issued. Based on this observation, we can define
the two metrics mentioned above. If an update operation is issued
at time $t$, then we note $V_{G}(t)$ the new value of the version
vector of the global state at that time, and $V_{i}(t)$ the version
vector of replica $i$ at time $t$ (while playing the application
scenario our simulator maintains a version vector for each replica).
We note $D_{i}(t)$ the convergence distance between $V_{i}(t)$ and
$V_{G}(t)$. It is simply the number of updates that are seen in $V_{G}(t)$
but not in $V_{i}(t)$, and it can be computed as $D_{i}(t)=\sum_{k=0}^{n-1}V_{G}(t)_{_{k}}-\sum_{k=0}^{n-1}V_{i}(t)_{_{k}}$.
(Note that the version vector of a replica cannot be higher than that
of the global state, since each replica's state is constantly lagging
behind the global state. So $D_{i}(t)$ is always positive.)

Computing the convergence latency is a bit trickiest, as it requires
logging all the values reached by version vectors during a simulation
run (that of the global state, but also that of each replica). Thus,
when the simulation log shows that the global state's version vector
has been updated at time $t$ and reached value $V_{G}(t)$, we must
parse the remainder of the simulation log and find for each replica
$i$ the earliest time $t'$ at which its local state has caught up
with $V_{G}(t)$, that is, $V_{i}(t')\geq V_{G}(t)$. Once $t'$ has
been found, the convergence latency for replica $i$ at time $t$
is simply defined as $L_{i}(t)=t'-t$. (Note that we do not simply
look for $V_{i}(t')=V_{G}(t)$, because as synchronizations go the
state of replica $i$ may ``jump over'' the exact state represented
by $V_{G}(t)$.) When an update operation is issued at time $t$ and
the global state's version vector becomes $V_{G}(t)$, $L_{i}(t)$
therefore shows how long it will take for replica $i$ to reach a
state that is equal to or after that of the current global state (i.e.,
how long it will take for replica $i$ to catch up with the current
global state).

Besides the convergence latency and convergence distance, we shall
also consider how costly it is for relays to maintain CRDT states
in a local store, and for replicas to synchronize with relays, rather
than just among replicas. While analyzing our simulation logs we will
therefore examine the number of states stored by relays in their local
store, as well as the number of states transmitted by replicas and
relays during synchronizations.

\subsection{Definition of simulation scenarios\label{subsec:Simulation-scenarios}}

Depending on the kinds of nodes considered (devices carried by pedestrians,
vehicles, drones, etc.), on their spatial density and radio range,
on their mobility patterns, on the nature and needs of the CRDT-based
distributed applications they may run, the variety of simulation scenarios
that may be considered is almost infinite. In this paper our goal
is not to explore the whole spectrum of possible scenarios. It is
simply to exemplify a few possible scenarios, while showing that assisting
CRDT replicas with relays in these scenarios can significantly improve
the synchronization of the replicas.

In the next sections we therefore explore three typical scenarios:
the first scenario involves a small set of mobile replicas assisted
by a large population of mobile relays (with churn), the second one
involves a small set of static replicas assisted by a larger population
of mobile relays, and the third one involves a large population of
mobile replicas assisted by a small set of static relays. 

For the sake of clarity, when referring to time-related measures in the
following, we will use sexagesimal-based notations for duration (e.g.,
$3'17''$ or $2\mathrm{h}52'$), and usual notations for dates, more
precisely relative times in an experiment (e.g., 13:05 for 13~hours
5~minutes after the beginning of the experiment).

\subsection{\label{subsec:Scenario_Ostermalm}\"Ostermalm scenario}

\begin{figure}[H]
\begin{centering}
\includegraphics[width=8.2cm]{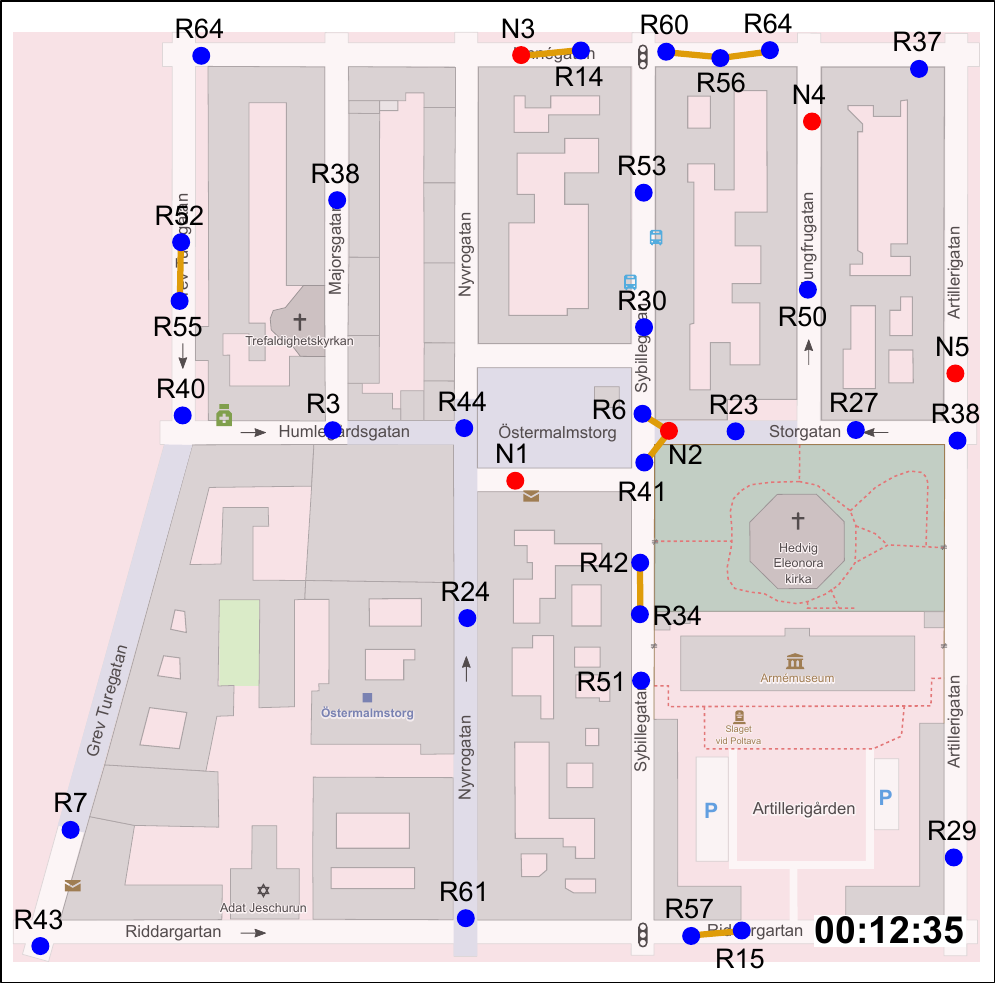}
\par\end{centering}
\caption{\label{fig:ostermalm_snapshot}Snapshot of the \"Ostermalm scenario
(with replica nodes in red and relay nodes in blue)}
\end{figure}

In this section we consider an opportunistic network whose mobile
nodes are devices carried by pedestrians. These pedestrians wander
in the streets of the \"Ostermalm district in downtown Stockholm. The
traceset describing their mobility is available in the CRAWDAD database
(now hosted by \emph{IEEE DataPort}\footnote{\url{https://ieee-dataport.org/collections/crawdad}})~\cite{comcom16pajevic,kth/walkers}.
The traceset we consider here is referred to as \emph{ostermalm\_sparse\_run1}
in that dataset. It covers an interval of 5 hours, during which a
continuous flow of pedestrians traverse the area, each traversal taking
only a few minutes (see details in Table~\ref{fig:ostermalm_details}).
Although the total number of pedestrians considered in this traceset
is large, the number of pedestrians that are present in the area at
the same time never exceeds a few dozens (maximal value: 57), as can
be observed in Fig.~\ref{fig:ostermalm_snapshot}, and also in a
video that is available on our website\footnote{\label{fn:ref_videos}\url{https://casa-irisa.univ-ubs.fr/lepton/videos.html}}.

This particular traceset is interesting because of the high churn
rate it exhibits. It makes it possible for us to illustrate a synchronization
scenario involving a small population \ensuremath{P} of nodes carrying
replicas, which move in the \"Ostermalm area and never leave this area,
while a far larger population of nodes \ensuremath{R} only traverse
the area but can yet serve as relays between the replicas. The \"Ostermalm
traceset mentioned above provides the mobility scenario for the population
of relays. For the replicas, we modelled the \"Ostermalm district as
a simple graph, each edge in the graph modelling a street, and we
used a graph walk model to define a mobility scenario for the mobile
nodes that are assumed to carry replicas in our experiment. We then
merged the two distinct mobility scenarios (replicas and relays),
and calculated the radio contacts between all nodes, assuming a transmission
range of 15 meters between any pair of nodes. The resulting contact
scenario is a time-varying graph that can be played by the parsers
defined in the GraphStream library, and thus by our simulator. 

Fig.~\ref{fig:ostermalm_snapshot} shows a snapshot of the simulation
area, with red nodes modelling replicas, and blue nodes modelling
relays. Orange edges depict radio contacts. In this figure it can
be observed that the network is indeed an OppNet, as each node has
at most a few neighbors, and the whole network is highly partitioned.

\begin{table}[h]
\centering{}\caption{\foreignlanguage{english}{\label{fig:ostermalm_details}Details about the \"Ostermalm scenario}}
\begin{tabular}{|c|c|}
\cline{2-2} 
\multicolumn{1}{c|}{} & {\rule{0pt}{2.5ex}\small{}\textbf{Values ({*} = min/max/avg/stdev)} \hspace*{-1ex}}\tabularnewline
\hline 
{\small{}\rule{0pt}{2.5ex}}\textbf{\small{}A: Mobility scenario for
relays} & {\small{}\"Ostermalm walkers~\cite{kth/walkers}}\tabularnewline
\hline 
{\small{}\rule{0pt}{2.5ex}Duration of the scenario} & {\small{}5\,h}\tabularnewline
\hline 
{\small{}\rule{0pt}{2.5ex}Nb of nodes} & {\small{}2092}\tabularnewline
\hline 
{\small{}\rule{0pt}{2.5ex}Traversal time per node} & {\small{}$7.8''\,/\,32'16''\,/\,5'25''\,/\thinspace4'09''$ }\textsuperscript{{\small{}({*})}}\tabularnewline
\hline 
{\small{}\rule{0pt}{2.5ex}Entry rate} & {\small{}0.01 node/s (Poisson)}\tabularnewline
\hline 
{\small{}\rule{0pt}{2.5ex}Speed} & {\small{}{[}0.6--2.0{]}\,m/s}\tabularnewline
\hline 
\hline 
{\small{}\hspace*{-1ex}\rule{0pt}{2.5ex}}\textbf{\small{}B: Mobility
scenario for replicas}{\small{}\hspace*{-1ex}} & {\small{}Graph walk}\tablefootnote{{\small{}Nodes walking on the streets of the \"Ostermalm district, without
ever leaving the district.}}\tabularnewline
\hline 
{\small{}\rule{0pt}{2.5ex}Nb of nodes} & {\small{}5}\tabularnewline
\hline 
{\small{}\rule{0pt}{2.5ex}Speed} & {\small{}{[}0.3--1.5{]}\,m/s}\tabularnewline
\hline 
{\small{}\rule{0pt}{2.5ex}Delay between flights} & {\small{}{[}10--60{]}\,s}\tabularnewline
\hline 
\hline 
\multirow{2}{*}{{\small{}\rule{0pt}{2.5ex}}\textbf{\small{}Contact scenario}} & {\small{}\rule{0pt}{2.5ex}Based on a combination}\tabularnewline
 & {\small{}of A and B}\tabularnewline
\hline 
{\small{}\rule{0pt}{2.5ex}Transmission range} & {\small{}15\,m}\tabularnewline
\hline 
{\small{}\rule{0pt}{2.5ex}Number of contacts} & \multirow{2}{*}{{\small{}$11\,793$}}\tabularnewline
{\small{}between pairs of nodes} & \tabularnewline
\hline 
{\small{}\rule{0pt}{2.5ex}Duration of contacts} & {\small{}$0.2''\thinspace/\thinspace5'48''\thinspace/\thinspace17''\thinspace/\thinspace22''$
}\textsuperscript{{\small{}({*})}}\tabularnewline
\hline 
\hline 
\multirow{2}{*}{{\small{}\rule{0pt}{2.5ex}}\textbf{\small{}Application scenario}} & {\small{}\rule{0pt}{2.5ex}1325 update operations}\tabularnewline
 & {\small{}between 00:05 and 04:30}\tabularnewline
\hline 
\end{tabular}
\end{table}

\subsubsection{Application scenario}

We defined a simple application scenario, whereby each node carrying
a replica issues an update operation on its local state once every
minute in the {[}00:05--04:30{]} interval, with a randomly selected
jitter at start time so that all replicas behave asynchronously. This
activity interval allows the nodes to populate the area at the beginning
of the simulation (warm-up period), and to propagate and synchronize
states for a while after the last update operation has been issued
on a replica (cool-down period). Overall, 1325 update operations are
issued on the 5 replicas (265 each) during the interval considered.

\subsubsection{Experimental procedure}

In order to show the impact of using mobile relays to help in the
synchronization of replicas, we ran the above-mentioned scenario (i.e.,
contact + application scenario), while varying the ratio of pedestrians
that served as relays for replica synchronization. For each simulation
run, pedestrians were simply dynamically assigned a role (either relay
or nothing) upon entering the area. For example, for ratio 33\%, one
pedestrian out of three was assigned the role of relay, while the
next two pedestrians entering the area were assigned no role in the
simulation. The number of replica nodes was maintained at 5 for all
simulation runs. The number of synchronizations observed during each
simulation run is shown in Table~\ref{fig:ostermalm_syncs}.

\begin{table}[h]
\centering{}\caption{\label{fig:ostermalm_syncs}Statistics about the number of synchronizations
observed against the ratio of pedestrians serving as relays (\"Ostermalm
scenario)}
\begin{tabular}{|l|c|c|c|c|c|c|}
\hline
{\small\rule{0pt}{2.5ex}Ratio of pedestrians serving as relays} & {\small 0\% (no relay)} & {\small 10\%} & {\small 20\%} & {\small 33\%} & {\small 50\%} & {\small 100\%}\tabularnewline
\hline
{\small\rule{0pt}{2.5ex}Number of sync. between pair of nodes} & {\small 111} & {\small 478} & {\small 930} & {\small 2008} & {\small 3775} & {\small\hspace*{-1ex}11793\hspace*{-1ex}}\tabularnewline
\hline
{\small\rule{0pt}{2.5ex}Number of sync. between pairs of replicas} & {\small 111} & {\small 111} & {\small 111} & {\small 111} & {\small 111} & {\small 111}\tabularnewline
\hline
{\small\rule{0pt}{2.5ex}Number of sync. between replicas and relays} & \multirow{1}{*}{{\small -}} & {\small 288} & {\small 531} & {\small 961} & {\small 1507} & {\small 2816}\tabularnewline
\hline
{\small\rule{0pt}{2.5ex}Number of sync. between pairs of relays} & \multirow{1}{*}{{\small -}} & {\small 79} & {\small 288} & {\small 936} & {\small 2157} & {\small 8866}\tabularnewline
\hline
\end{tabular}
\end{table}

\subsubsection{Simulation results}

\begin{figure}[H]
\begin{centering}
\includegraphics[width=17.3cm]{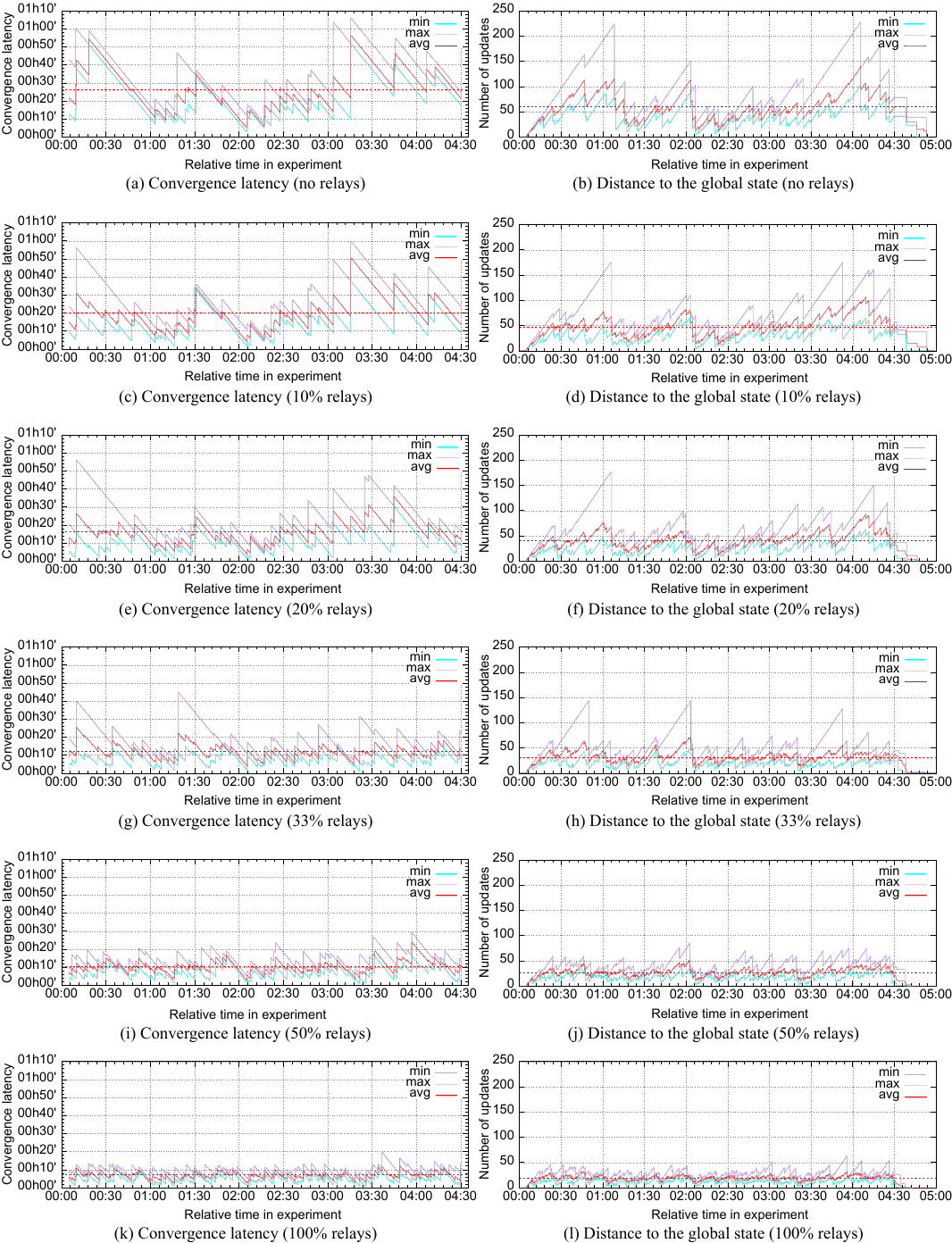}
\end{centering}
\caption{\label{fig:ostermalm1}Evolution of the convergence
latency (left) and distance to the global state (right) when running
the \"Ostermalm scenario with different ratios of relays}
\end{figure}

Fig.~\ref{fig:ostermalm1} shows the evolution of
the convergence latency and convergence distance when running the
\"Ostermalm scenario with different ratios of relays (from 0\% to 100\%).
In this figure the left column shows the latency, and the right column
shows the distance. The same vertical scale is used in all subfigures
in the left column (so they can be compared at a glance), and the
same is true in the right column. Each subfigure shows the evolution
of the minimal, maximal, and average latency (resp. distance) for
the population of 5 replicas considered in this scenario. The dashed
horizontal line shows the average value of all average latencies (resp.
distances) over time. 

Let us first consider the case where the scenario is run with no relays
(Fig.~\ref{fig:ostermalm1}{\textcolor{blue}a} and~\ref{fig:ostermalm1}{\textcolor{blue}b}).
In that case, the 5 replicas can only synchronize when getting in
direct contact with one another, which only occurs 111 times in the
5 hour interval, as shown in Table~\ref{fig:ostermalm_syncs}. The
convergence latency observed on all replicas is thus quite high: it
revolves around 26 minutes on average (horizontal dashed line in~Fig.~\ref{fig:ostermalm1}{\textcolor{blue}a}).
As for the average distance to the global state, it revolves around
60 update operations. On average a replica is thus 26 minutes late,
and 60 update operations late, with respect to the global state. Note
that the three curves in Fig.~\ref{fig:ostermalm1}{\textcolor{blue}a}
stop abruptly at 04:30. This is normal, because this figure shows
a value (either min, max or average latency) whenever the global state
changes, that is, whenever an update operation is issued on a replica.
Since the last update operation occurs at 04:30 in the application
scenario considered here, no latency value can be computed after that
last event. In contrast the evolution of the convergence distance
can still be observed after the last update operation, as shown in
Fig.~\ref{fig:ostermalm1}{\textcolor{blue}b}. In this figure the minimal,
maximal, and average distances all reach 0 eventually, which confirms
that all replicas eventually catch up with the global state before
the end of the simulation period.

Running the simulation with no relays is clearly the worst-case scenario,
and it is meant to serve as a baseline for our analysis. Let us now
consider the best-case scenario, which consists in enrolling all the
pedestrians that traverse the area as relays (Fig.~\ref{fig:ostermalm1}{\textcolor{blue}k}
and~\ref{fig:ostermalm1}{\textcolor{blue}l}). In that case, the 5 replicas
can still synchronize directly with one another, but they can also
synchronize with relays (which occurs 2816 times in the 5 hour interval:
see Table~\ref{fig:ostermalm_syncs}), and relays can synchronize
together (which occurs 8866 times). The convergence latency and convergence
distance are then reduced significantly: the average latency over
time is $7'10''$, and the average distance is only 18 update operations.

This comparison of the worst-case and best-case scenarios confirms
that using relays for synchronization can significantly reduce the
time it takes for replicas to catch up with the global state, and
it also reduces the distance between a replica's state and the global
state (in terms of update operations missing on the replica). 

\begin{figure}[H]
\begin{centering}
\includegraphics[width=17.5cm]{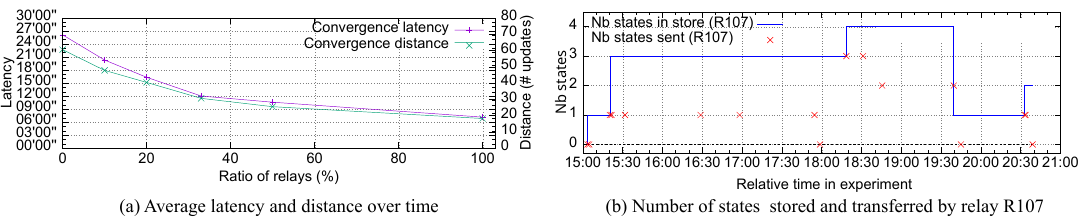}
\end{centering}
\caption{\label{fig:ostermalm2}Evolution of the average latency
and distance over time against the ratio of relays (left), and evolution
of the number of states stored and transferred by a single relay (right) while running the \"Ostermalm scenario}
\end{figure}

When only a subpopulation of the pedestrians that traverse the \"Ostermalm
area serve as relays, the latency and distance of course fall between
the values observed for the worst-case and best-case scenarios, as
shown in Fig.~\ref{fig:ostermalm1}, and as synthesized
in Fig.~\ref{fig:ostermalm2}{\textcolor{blue}a}. Interestingly, the
latter figure shows that in the \"Ostermalm scenario considered here,
maximizing the number of relays brings little benefit in terms of
latency and distance reduction. Using 50\% relays is ``almost as
good'' as using 100\% relays. Based on this observation, it may be
tempting to try to find the optimal ratio of relays ---or even better,
the most effective relays---, considering what is deemed tolerable
in terms of convergence latency and distance. This kind of investigation
clearly falls out of the scope of this paper, though, as our main
objective is to demonstrate that relays can help replicas to synchronize.
Besides, no two OppNet mobility and/or contacts scenarios are similar,
so devising heuristics to adjust the number of relays in a scenario
(or select the nodes that should serve as relays) may be a daunting
task in many cases.

After considering how the use of relays can impact the convergence
latency and convergence distance, let us now examine the cost of using
relays to help in the synchronization of replicas. The synchronization
protocols described in Section~\ref{sec:Synchronization_algorithms}
have been designed so as to minimize the number of states stored by
relays, and also the number of states exchanged by neighbor nodes
when these nodes synchronize. Fig.~\ref{fig:ostermalm2}{\textcolor{blue}b}
shows the evolution of the number of states maintained by a randomly
selected relay in its local store at runtime, and the number of states
transferred by this relay when synchronizing with other nodes (either
replicas or relays). In this figure we deliberately focus on a single
relay, namely node R107, because considering many relays or even all
relays in the same figure would be too messy. In the figure, the evolution
of the number of states maintained in the relay's local store is shown
as a continuous blue line, and the number of states transferred by
the relay to a neighbor node (either a replica or a relay) is shown
as a red \textsf{X} symbol. Note that it takes only a few minutes
for R107 to traverse the \"Ostermalm area, and thus assist in the synchronization
of replicas.
Although there are 5 replicas in the scenario considered,
and the store of R107 may thus contain up to 5 conflicting states
simultaneously, the number of states actually stored by R107 never
exceeds 4, and falls down to 1 at 19:40, as R107 gets a chance to synchronize
with a replica. Moreover, when synchronizing with a neighbor node
(either relay or replica), R107 rarely has to send all the states
it owns to this neighbor. During a synchronization the number of states
sent by R107 to a neighbor is usually lower than the number of states
it owns (i.e., the red symbols are below the blue line most of the
time). These observations confirm that our synchronization protocols
are effective at minimizing both the amount of data stored by a relay,
and the amount of data it transfers to other nodes during synchronizations.

These observations about what happens on relay R107 can be generalized
to all other relays.  Table~\ref{fig:ostermalm_relay_stored}
shows statistics about the number of states maintained by the relays
in their store while running the \"Ostermalm scenario with 100\% relays and 50\%
relays\footnote{Similar results could be presented for lower ratios of
relays. They are omitted here for the sake of conciseness.}. The
figures presented in this table have been obtained by considering the
number of states maintained in each relay's store in between two
synchronizations involving this relay. Although a relay may have to
store up to 5 conflicting states in its store (because the scenario
considered involves 5 replicas), the number of states maintained in
the stores is actually lower most of the time.

\begin{table*}[h]
\centering{}\caption{\label{fig:ostermalm_relay_stored}Statistics about the number of states  maintained in stores
 during synchronizations (\"Ostermalm scenario with 100\% and 50\% relays)}
{\small{}}%
\begin{tabular}{|c|c|c|c|c|c|c|c|c|}
\cline{2-9} \cline{3-9} \cline{4-9} \cline{5-9} \cline{6-9} \cline{7-9} \cline{8-9} \cline{9-9} 
\multicolumn{1}{c|}{} & {\small{}\rule{0pt}{2.5ex}Nb of states in stores} & {\small{}0} & {\small{}1} & {\small{}2} & {\small{}3} & {\small{}4} & {\small{}5} & {\small{}Total}\tabularnewline
\hline 
\multirow{2}{*}{{\small{}100\% relays}} & {\small{}\rule{0pt}{2.5ex}Nb. of occurrences} & {\small{}2937} & {\small{}16297} & {\small{}10988} & {\small{}5372} & {\small{}1682} & {\small{}451} & {\small{}37321}\tabularnewline
\cline{2-9} \cline{3-9} \cline{4-9} \cline{5-9} \cline{6-9} \cline{7-9} \cline{8-9} \cline{9-9} 
 & {\small{}\rule{0pt}{2.5ex}Ratio} & {\small{}7.8\%} & {\small{}43.6\%} & {\small{}29.4\%} & {\small{}14.3\%} & {\small{}4.5\%} & {\small{}0.1\%} & {\small{}100\%}\tabularnewline
\hline 
\hline 
\multirow{2}{*}{{\small{}50\% relays}} & {\small{}\rule{0pt}{2.5ex}Nb. of occurrences} & {\small{}1281} & {\small{}5267} & {\small{}2143} & {\small{}667} & {\small{}278} & {\small{}-} & {\small{}9636}\tabularnewline
\cline{2-9} \cline{3-9} \cline{4-9} \cline{5-9} \cline{6-9} \cline{7-9} \cline{8-9} \cline{9-9} 
 & {\small{}\rule{0pt}{2.5ex}Ratio} & {\small{}13.2\%} & {\small{}54.6\%} & {\small{}22.2\%} & {\small{}6.9\%} & {\small{}2.8\%} & {\small{}-} & {\small{}100\%}\tabularnewline
\hline 
\end{tabular}
\end{table*}

Table~\ref{fig:ostermalm_relay_tsfers} shows statistics about the
number of states transferred by the relays during
synchronizations. During most synchronizations, a relay node only has
to transmit a few states to the neighbor node, and in many cases it
does not have to transmit any state at all (either because both nodes
are already synchronized, or because there is nothing the local relay
could send to the other node that would inflate the remote state).

\begin{table*}[h]
\centering{}\caption{\label{fig:ostermalm_relay_tsfers}Statistics about the number of
states transferred by relays during synchronizations (\"Ostermalm scenario
with 100\% and 50\% relays)}
{\small{}}%
\begin{tabular}{|c|c|c|c|c|c|c|c|c|}
\cline{2-9} \cline{3-9} \cline{4-9} \cline{5-9} \cline{6-9} \cline{7-9} \cline{8-9} \cline{9-9} 
\multicolumn{1}{c|}{} & {\small{}\rule{0pt}{2.5ex}Nb of states transferred per sync.} & {\small{}0} & {\small{}1} & {\small{}2} & {\small{}3} & {\small{}4} & {\small{}5} & {\small{}Total}\tabularnewline
\hline 
\multirow{2}{*}{{\small{}100\% relays}} & {\small{}\rule{0pt}{2.5ex}Nb. of occurrences} & {\small{}11517} & {\small{}5985} & {\small{}1 477} & {\small{}481} & {\small{}128} & {\small{}1} & {\small{}19589}\tabularnewline
\cline{2-9} \cline{3-9} \cline{4-9} \cline{5-9} \cline{6-9} \cline{7-9} \cline{8-9} \cline{9-9} 
 & {\small{}\rule{0pt}{2.5ex}Ratio} & {\small{}58.7\%} & {\small{}30.5\%} & {\small{}7.5\%} & {\small{}2.4\%} & {\small{}0.6\%} & {\small{}0.005\%} & {\small{}100\%}\tabularnewline
\hline 
\hline 
\multirow{2}{*}{{\small{}50\% relays}} & {\small{}\rule{0pt}{2.5ex}Nb. of occurrences} & {\small{}2984} & {\small{}1891} & {\small{}345} & {\small{}75} & {\small{}27} & {\small{}-} & {\small{}5322}\tabularnewline
\cline{2-9} \cline{3-9} \cline{4-9} \cline{5-9} \cline{6-9} \cline{7-9} \cline{8-9} \cline{9-9} 
 & {\small{}\rule{0pt}{2.5ex}Ratio} & {\small{}56.0\%} & {\small{}35.5\%} & {\small{}6.4\%} & {\small{}1.4\%} & {\small{}0.5\%} & {\small{}-} & {\small{}100\%}\tabularnewline
\hline 
\end{tabular}
\end{table*}

Table~\ref{fig:ostermalm_replica_tsfers} shows similar transmission
statistics, but for the replicas. The same observations as above can
be made, except that the number of states transferred by a replica to
another node never exceeds 1, since a replica only maintains one state
locally.

\begin{table}[h]
\centering{}\caption{\label{fig:ostermalm_replica_tsfers}Statistics about the number of
states transferred by replicas during synchronizations (\"Ostermalm
scenario with 100\% and 50\% relays)}
{\small{}}%
\begin{tabular}{|c|c|c|c|c|}
\cline{2-5} \cline{3-5} \cline{4-5} \cline{5-5} 
\multicolumn{1}{c|}{} & {\small{}\rule{0pt}{2.5ex}Nb of states transferred} & \multirow{2}{*}{{\small{}0}} & \multirow{2}{*}{{\small{}1}} & \multirow{2}{*}{{\small{}Total}}\tabularnewline
\multicolumn{1}{c|}{} & {\small{}per sync.} &  &  & \tabularnewline
\hline 
\multirow{2}{*}{{\small{}100\% relays}} & {\small{}\rule{0pt}{2.5ex}Nb. of occurrences} & {\small{}66} & {\small{}2013} & {\small{}2079}\tabularnewline
\cline{2-5} \cline{3-5} \cline{4-5} \cline{5-5} 
 & {\small{}\rule{0pt}{2.5ex}Ratio} & {\small{}3.2\%} & {\small{}96.8\%} & {\small{}100\%}\tabularnewline
\hline 
\hline 
\multirow{2}{*}{{\small{}50\% relays}} & {\small{}\rule{0pt}{2.5ex}Nb. of occurrences} & {\small{}51} & {\small{}1179} & {\small{}1230}\tabularnewline
\cline{2-5} \cline{3-5} \cline{4-5} \cline{5-5} 
 & {\small{}\rule{0pt}{2.5ex}Ratio} & {\small{}4.1\%} & {\small{}95.9\%} & {\small{}100\%}\tabularnewline
\hline 
\end{tabular}
\end{table}

\subsection{\label{subsec:Scenario_VBN}VBN scenario}

With the \"Ostermalm scenario we have considered a scenario with a small
population of mobile replicas, and a large population of mobile relays,
with a lot of churn in the latter population. We will now consider
a scenario involving a small population of static replicas, and a
larger yet fixed population of mobile relays (i.e., no churn). 

To build this scenario we rely on a dataset describing the mobility
of buses serving the city of Vannes (France)~\cite{ubs/vbn}. This
dataset involves 60 buses, running over 10 bus lines, and running
for about 14 hours every day. These buses will be considered as mobile
relays in our scenario. To these 60 relays we add 10 nodes that we
place at different locations (typically, buildings offering public
services) all over the city. These 10 nodes are static, and will be
considered as replicas in our scenario. 
\begin{figure}[h]
\begin{centering}
\includegraphics[width=12cm]{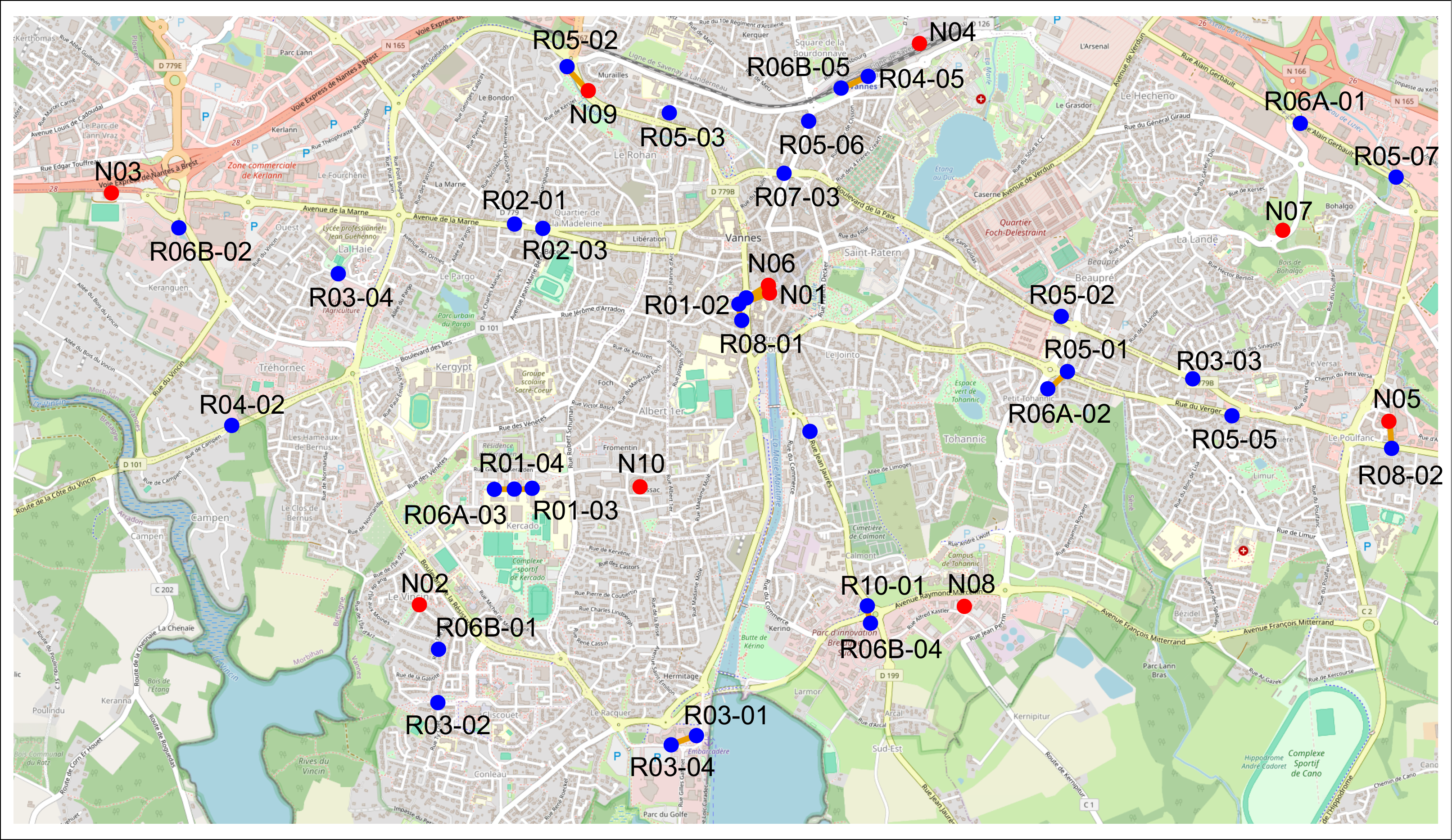}
\par\end{centering}
\caption{\label{fig:vbn_snapshot}Snapshot of the VBN scenario, with mobile
relay nodes (buses) in blue and static replica nodes in red}
\end{figure}

\begin{table}[h]
\centering{}\caption{\label{fig:vbn_details}Details about the VBN scenario}
\begin{tabular}{|c|c|}
\cline{2-2} 
\multicolumn{1}{c|}{} & {\small{}\hspace*{-1ex}\rule{0pt}{2.5ex}\textbf{Values ({*} = min/max/avg/stdev)}\hspace*{-1ex}}\tabularnewline
\hline 
{\small{}\hspace*{-1ex}\rule{0pt}{2.5ex}}\textbf{\small{}A:
Mobility scenario for relays}{\small{}\hspace*{-1ex}} & {\small{}Vannes Bus Network~\cite{ppna22guidec}}\tabularnewline
\hline 
{\small{}\rule{0pt}{2.5ex}Duration of the scenario} & {\small{}14h$14'$ (from 06:25 to 20:39)}\tabularnewline
\hline 
{\small{}\rule{0pt}{2.5ex}Nb of nodes} & {\small{}60}\tabularnewline
\hline 
\hline 
\multirow{2}{*}{\textbf{\small{}B: Static scenario for replicas}} & {\small{}\rule{0pt}{2.5ex}Nodes placed wide apart}\tabularnewline
 & {\small{}at static locations}\tabularnewline
\hline 
{\small{}\rule{0pt}{2.5ex}Nb of nodes} & {\small{}10}\tabularnewline
\hline 
\hline 
\multirow{2}{*}{\textbf{\small{}Contact scenario}} & {\small{}\rule{0pt}{2.5ex}Based on a combination}\tabularnewline
 & {\small{}of A and B}\tabularnewline
\hline 
{\small{}\rule{0pt}{2.5ex}Transmission range} & {\small{}200 m}\tabularnewline
\hline 
{\small{}\rule{0pt}{2.5ex}Number of contacts} & \multirow{2}{*}{{\small{}10731}}\tabularnewline
{\small{}between pairs of nodes} & \tabularnewline
\hline 
{\small{}\rule{0pt}{2.5ex}Duration of contacts} & {\small{}$1''\,/\,1\mathrm{h}20'\,/\,2'23''\,/\,12'39''$ }\textsuperscript{{\small{}({*})}}\tabularnewline
\hline 
\hline 
\multirow{2}{*}{\textbf{\small{}Application scenario}} & {\small{}\rule{0pt}{2.5ex}1320 update operations}\tabularnewline
 & {\small{}between 01:00 and 12:05}\tabularnewline
\hline 
\end{tabular}
\end{table}

By combining the mobility of the 60 buses (relays) with the presence
of 10 additional static nodes (replicas), and assuming a transmission
range of 200 meters, we obtain a single scenario involving 10731 contacts
between pairs of nodes (either replicas or relays, see details in
Table~\ref{fig:vbn_details}). Note that in this scenario the replicas
have deliberately been placed wide apart (see the snapshot in Fig.~\ref{fig:vbn_snapshot}),
so they cannot synchronize directly with one another. The only way
for these replicas to synchronize is to rely on the assistance provided
by the mobile relays (buses). Fig.~\ref{fig:vbn_snapshot} only shows
part of the city, so some of the nodes are not displayed.

\subsubsection{Application scenario}

A simple application scenario is defined, whereby each node carrying
a replica issues an update operation on its local state once every
5 minutes in the {[}01:00{\small{}--}12:00{]} interval (relative
time), with a randomly selected jitter at start time so that all replicas
behave asynchronously. Overall, 1320 update operations are issued
on the 10 replicas (132 each) during the interval considered. In Fig.~\ref{fig:vbn_snapshot},
replicas are depicted with red nodes and relays with blue nodes. 

\subsubsection{Simulation results}
\begin{figure}[h]
\begin{centering}
\includegraphics[width=16.5cm]{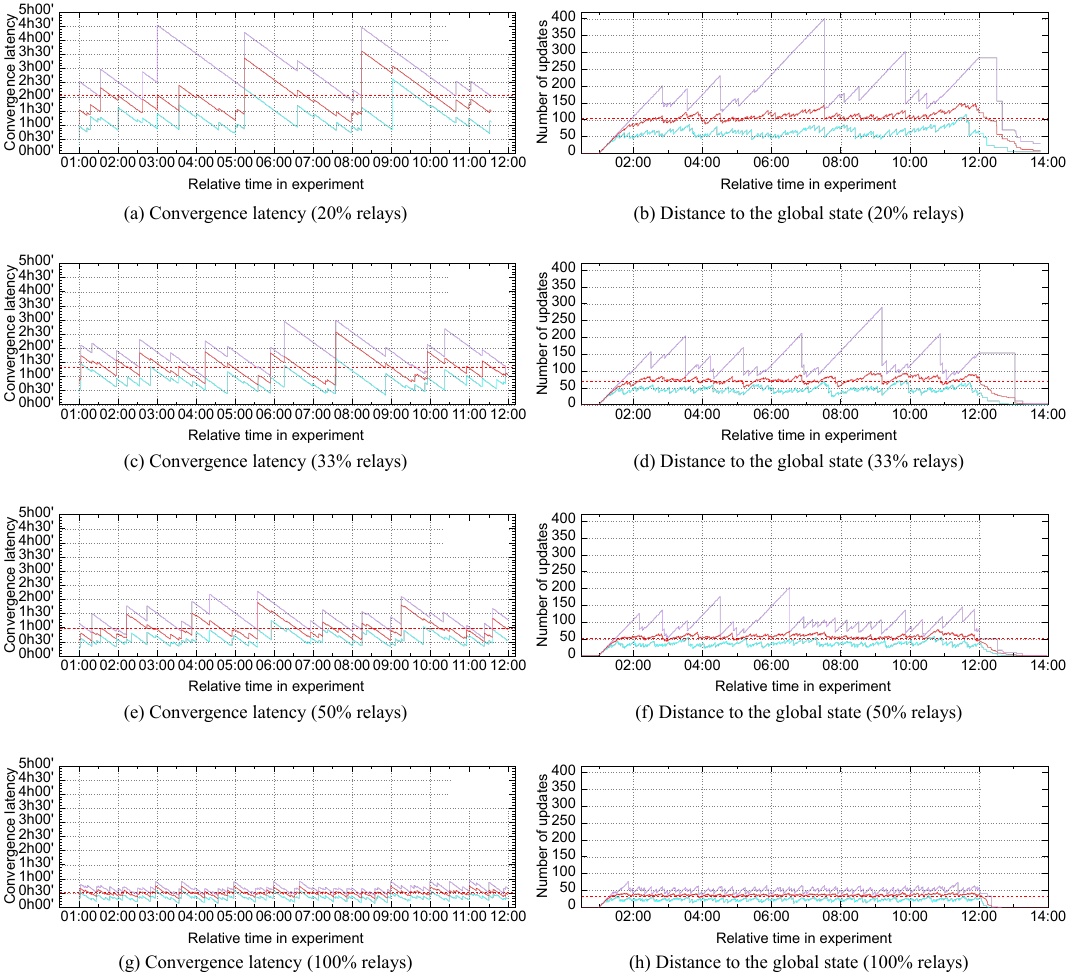}
\end{centering}
\caption{\label{fig:kiceo1}Evolution of the convergence latency
(left) and distance to the global state (right) when running the VBN
scenario with different ratios of relays}
\end{figure}

\begin{figure}[h]
\begin{centering}
\includegraphics[width=16.5cm]{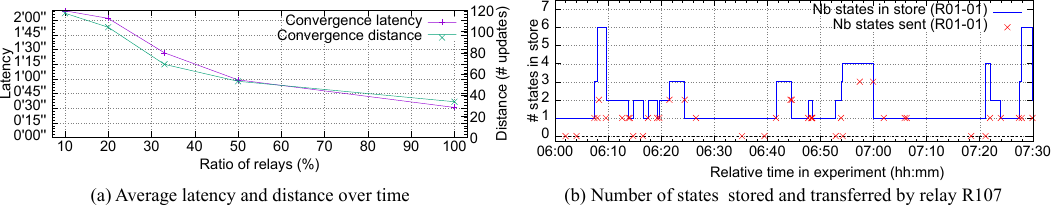}
\end{centering}
\caption{\label{fig:kiceo2}Evolution of the average latency
and distance over time against the ratio of relays (left), and evolution
of the number of states stored and transferred by a single relay (right)
while running the VBN scenario}
\end{figure}

Fig.~\ref{fig:kiceo1} and~\ref{fig:kiceo2}
present results that have been obtained with the VBN scenario, using
the same experimental procedure and the same metrics as for the \"Ostermalm
scenario. Note that this time we did not run the scenario with 0\%
relays, since in this scenario the replicas are static and would be
unable to synchronize without the assistance of mobile relays. Besides,
we did not decrease the ratio of relays below 20\%, because under
that ratio some of the bus lines do not include buses serving as relays
anymore, so some of the static replicas are not visited by relays
anymore, and therefore could not synchronize in such conditions. In
fact it can be observed in Fig.~\ref{fig:kiceo1}{\textcolor{blue}b} that
the minimal convergence distance reaches 0 at the end of the simulation
period, but this is not the case for the maximal and average distances.
This is an indication that some of the replicas fail to ``catch up''
with the global state when only 20\% relays are used in this scenario
(presumably they would catch up if the cool-down interval was longer
after the last update operation of a replica). The same observation
can be made in Fig.~\ref{fig:kiceo1}{\textcolor{blue}d}, although it is
less flagrant in that figure. In contrast all replicas manage to converge
when 50\% or 100\% relays are involved in the simulation.

In Fig.~\ref{fig:kiceo1} it can be observed that
the average latency over time is about 30 minutes (with an average
distance of 33 update operations) when all buses serve as relays,
and these figures degrade quite rapidly when only some of the buses
are used as relays. This observation is confirmed in Fig.~\ref{fig:kiceo2}{\textcolor{blue}a}.

Fig.~\ref{fig:kiceo2}{\textcolor{blue}b} shows the evolution of the
number of states maintained by relay R01-01 (i.e., bus number 01 in
bus line number 01) in its local store at runtime, and the number of
states transferred by this relay when synchronizing with neighbor
nodes.  In this figure we focus on a 90 minute interval (instead of
the whole 14 hours) to ensure that the figure is readable. Although
the number of replicas in this scenario is 10, and the relay's store
may thus occasionally contain up to 10 conflicting states, we can
observe that the relay's store actually contains much less states most
of the time.  Besides, the number of states sent to other nodes during
synchronizations is usually even lower.

Statistics about the number of synchronizations are presented in
Table~\ref{fig:kiceo_syncs}. Statistics about the number of states
maintained by the relays in their store are presented in Table~\ref{fig:kiceo_relay_stored},
and statistics about the number of states sent by the replicas and
relays during synchronizations are available in
Tables~\ref{fig:kiceo_relay_tsfers} and~\ref{fig:kiceo_replica_tsfers}.
It can again be observed that the
number of conflicting states a relay has to maintain in its store is
usually significantly lower than the theoretical max (i.e., the number
of replicas). Besides, when synchronizing with another node a relay
usually has to transmit only a couple of states.

\begin{table}[h]
\centering{}\caption{\label{fig:kiceo_syncs}Statistics about the number of synchronizations
observed against the ratio of buses serving as relays (VBN scenario)}
\begin{tabular}{|c|c|c|c|c|}
\hline 
{\small{}\rule{0pt}{2.5ex}Ratio of buses serving as relays} & {\small{}20\% (12 buses)} & {\small{}33\% (20 buses)} & {\small{}50\% (30 buses)} & {\small{}100\% (60 buses)}\tabularnewline
\hline 
{\small{}\rule{0pt}{2.5ex}Number of sync. between pairs of
nodes} & {\small{}864} & {\small{}1816} & {\small{}3692} & {\small{}10731}\tabularnewline
\hline 
{\small{}\rule{0pt}{2.5ex}Number of sync. between replicas
and relays} & {\small{}473} & {\small{}909} & {\small{}1 285} & {\small{}2339}\tabularnewline
\hline 
{\small{}\rule{0pt}{2.5ex}Number of sync. between pairs of
relays} & {\small{}391} & {\small{}907} & {\small{}2407} & {\small{}8392}\tabularnewline
\hline 
\end{tabular}
\end{table}

\begin{table}[h]
\centering{}\caption{\label{fig:kiceo_relay_stored}Statistics about the number of states
maintained in stores by relays during synchronizations (VBN scenario)}
\begin{tabular}{|c|c|c|c|c|c|c|c|c|c|c|c|c|}
\cline{2-13}
\multicolumn{1}{c|}{} & {\small Nb. of states} & \multirow{2}{*}{{\small 0}} & \multirow{2}{*}{{\small 1}} & \multirow{2}{*}{{\small 2}} & \multirow{2}{*}{{\small 3}} & \multirow{2}{*}{{\small 4}} & \multirow{2}{*}{{\small 5}} & \multirow{2}{*}{{\small 6}} & \multirow{2}{*}{{\small 7}} & \multirow{2}{*}{{\small 8}} & \multirow{2}{*}{{\small 9}} & \multirow{2}{*}{{\small Total}}\tabularnewline
\multicolumn{1}{c|}{} & {\small in stores} &  &  &  &  &  &  &  &  &  &  & \tabularnewline
\hline
{\small \rule{0pt}{2.5ex}100\%} & {\small Nb. of occurrences} & {\small 120} & {\small 17547} & {\small 7021} & {\small 5896} & {\small 3162} & {\small 985} & {\small 460} & {\small 242} & {\small 53} & {\small 8} & {\small 35494}\tabularnewline
\cline{2-13}
{\small relays} & {\small Ratio} & {\small 0.3\%} & {\small 49.4\%} & {\small 19.7\%} & {\small 16.6\%} & {\small 8.9\%} & {\small 2.7\%} & {\small 1.2\%} & {\small 0.6\%} & {\small 0.1\%} & {\small 0.02\%} & {\small 100\%}\tabularnewline
\hline
\hline
{\small \rule{0pt}{2.5ex}50\%} & {\small Nb. of occurrences} & {\small 26} & {\small 5994} & {\small 2480} & {\small 1656} & {\small 506} & {\small 97} & {\small 10} & {\small -} & {\small -} & {\small -} & {\small 10769}\tabularnewline
\cline{2-13}
{\small relays} & {\small Ratio} & {\small 0.2\%} & {\small 55.6\%} & {\small 23.0\%} & {\small 15.3\%} & {\small 4.6\%} & {\small 0.9\%} & {\small 0.09\%} & {\small -} & {\small -} & {\small -} & {\small 100\%}\tabularnewline
\hline
\end{tabular}
\end{table}

\begin{table}[h]
\centering{}\caption{\label{fig:kiceo_relay_tsfers}Statistics about the number of states
transferred by relays during synchronizations (VBN scenario)}
\begin{tabular}{|c|c|c|c|c|c|c|c|c|c|c|c|c|c|}
\cline{2-14} \cline{3-14} \cline{4-14} \cline{5-14} \cline{6-14} \cline{7-14} \cline{8-14} \cline{9-14} \cline{10-14} \cline{11-14} \cline{12-14} \cline{13-14} \cline{14-14} 
\multicolumn{1}{c|}{} & {\small{}\rule{0pt}{2.5ex}Nb of states} & \multirow{2}{*}{{\small{}0}} & \multirow{2}{*}{{\small{}1}} & \multirow{2}{*}{{\small{}2}} & \multirow{2}{*}{{\small{}3}} & \multirow{2}{*}{{\small{}4}} & \multirow{2}{*}{{\small{}5}} & \multirow{2}{*}{{\small{}6}} & \multirow{2}{*}{{\small{}7}} & \multirow{2}{*}{{\small{}8}} & \multirow{2}{*}{{\small{}9}} & \multirow{2}{*}{{\small{}10}} & \multirow{2}{*}{{\small{}Total}}\tabularnewline
\multicolumn{1}{c|}{} & {\small{}\hspace*{-1ex}transferred per sync.} &  &  &  &  &  &  &  &  &  &  &  & \tabularnewline
\hline 
{\small{}100\%} & {\small{}\hspace*{-1ex}\rule{0pt}{2.5ex}Nb. of occurrences} & {\small{}10000} & {\small{}6121} & {\small{}1582} & {\small{}650} & {\small{}201} & {\small{}94} & {\small{}47} & {\small{}13} & {\small{}2} & {\small{}-} & {\small{}-} & {\small{}18710}\tabularnewline
\cline{2-14} \cline{3-14} \cline{4-14} \cline{5-14} \cline{6-14} \cline{7-14} \cline{8-14} \cline{9-14} \cline{10-14} \cline{11-14} \cline{12-14} \cline{13-14} \cline{14-14} 
{\small{}\hspace*{-1ex}relays} & {\small{}\rule{0pt}{2.5ex}Ratio} & {\small{}53.4\%} & {\small{}32.7\%} & {\small{}8.4\%} & {\small{}3.4\%} & {\small{}1.0\%} & {\small{}0.5\%} & {\small{}0.2\%} & {\small{}0.07\%} & {\small{}0.01\%} & {\small{}-} & {\small{}-} & {\small{}100\%}\tabularnewline
\hline 
\hline 
{\small{}50\%} & {\small{}\hspace*{-1ex}\rule{0pt}{2.5ex}Nb. of occurrences} & {\small{}2 669} & {\small{}2 535} & {\small{}571} & {\small{}139} & {\small{}38} & {\small{}3} & {\small{}-} & {\small{}-} & {\small{}-} & {\small{}-} & {\small{}-} & {\small{}5955}\tabularnewline
\cline{2-14} \cline{3-14} \cline{4-14} \cline{5-14} \cline{6-14} \cline{7-14} \cline{8-14} \cline{9-14} \cline{10-14} \cline{11-14} \cline{12-14} \cline{13-14} \cline{14-14} 
{\small{}\hspace*{-1ex}relays} & {\small{}\rule{0pt}{2.5ex}Ratio} & {\small{}44.8\%} & {\small{}42.5\%} & {\small{}9.5\%} & {\small{}2.3\%} & {\small{}0.6\%} & {\small{}0.05\%} & {\small{}-} & {\small{}-} & {\small{}-} & {\small{}-} & {\small{}-} & {\small{}100\%}\tabularnewline
\hline 
\end{tabular}
\end{table}

\begin{table}[h]
\centering{}\caption{\label{fig:kiceo_replica_tsfers}Statistics about the number of states
transferred by replicas during synchronizations (VBN scenario)}
\begin{tabular}{|c|c|c|c|c|}
\cline{2-5} \cline{3-5} \cline{4-5} \cline{5-5}
\multicolumn{1}{c|}{} & {\small{}\rule{0pt}{2.5ex}Nb of states transferred} & \multirow{2}{*}{{\small{}0}} & \multirow{2}{*}{{\small{}1}} & \multirow{2}{*}{{\small{}Total}}\tabularnewline
\multicolumn{1}{c|}{} & {\small{}per sync} &  &  & \tabularnewline
\hline
{\small{}100\%} & {\small{}\rule{0pt}{2.5ex}Nb. of occurrences} & {\small{}413} & {\small{}1926} & {\small{}2339}\tabularnewline
\cline{2-5} \cline{3-5} \cline{4-5} \cline{5-5}
{\small{}relays} & {\small{}\rule{0pt}{2.5ex}Ratio} & {\small{}17.6\%} & {\small{}82.4\%} & {\small{}100\%}\tabularnewline
\hline
\hline
{\small{}50\%} & {\small{}\rule{0pt}{2.5ex}Nb. of occurrences} & {\small{}0} & {\small{}1141} & {\small{}1141}\tabularnewline
\cline{2-5} \cline{3-5} \cline{4-5} \cline{5-5}
{\small{}relays} & {\small{}\rule{0pt}{2.5ex}Ratio} & {\small{}0.0\%} & {\small{}100.0\%} & {\small{}100\%}\tabularnewline
\hline
\end{tabular}
\end{table}

\clearpage

\subsection{\label{subsec:Scenario_Disaster}Disaster relief scenario}

With both the \"Ostermalm scenario and the VBN scenario we have
considered a small population of replicas, assisted by a large
population of relays. We will now consider the opposite, that is, a
scenario involving a large population of replicas (so the scalability
of our approach can be verified), assisted by a small population of
relays.

None of the datasets available in the CRAWDAD database would fit our
needs to design such a scenario, so we resorted to building a purely
synthetic one. This synthetic scenario is meant to illustrate what
could be observed in a disaster relief situation, when the local
telecommunication infrastructure is inoperative and road vehicles are
hardly usable. A population of 1000 rescue workers (medical staff,
firefighters, field service technicians, etc.) is assumed to be spread
over a 10\,$\times$10~km${{}^2}$ area, moving only on foot because
road vehicles cannot be used. These workers use handheld devices
running a CRDT-based application to share information, such as their
own location, as well as the locations of points of interest (field
hospitals, medical and technical supplies, etc.). Because of the high
amount of obstacles at ground level, the radio range for
device-to-device transmissions is assumed to be only 50
meters. However, a flotilla of drones roams the area. These drones
move from place to place to deliver medical supplies where needed, but
while flying they also serve as fast-moving relays to disseminate
information among all rescue workers. As the drones have a better line
of sight than ground-based devices, the radio range for drone-to-drone
and drone-to-ground transmissions is assumed to be 200 meters.

The Levy Walk mobility model, which mimicks quite accurately the
mobility of human beings~\cite{ton11rhee}, is used for rescue
workers. The Random Waypoint mobility model is used for the drones,
which as explained above roam the area with no pre-defined flight plan
but according to needs. All simulation parameters are detailed in
Table~\ref{fig:disaster_details}.

\begin{figure}[H]
\begin{centering}
\includegraphics[width=8.2cm]{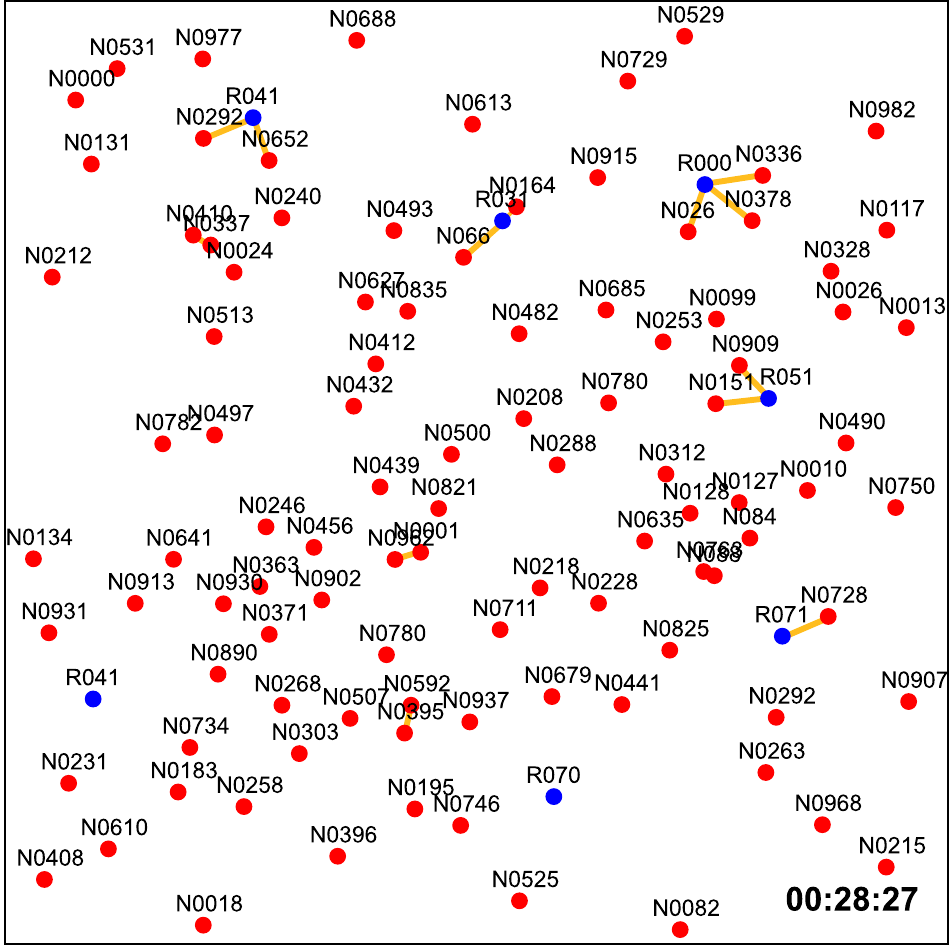}
\par\end{centering}
\caption{\label{fig:disaster_snapshot}Snapshot of the Disaster Relief scenario,
with replica nodes (rescue workers) in red and relay nodes (drones) in blue}
\end{figure}

By combining the mobility of the 1000 replicas (rescue workers) with
that of the 100 relays (drones), assuming a transmission range of 50
meters for ground-to-ground transmissions and 200 meters for
air-to-ground and air-to-air transmissions, we obtain a single
scenario involving 439552 contacts between pairs of nodes (either
replicas or relays) over 24 hours. Note that because of the reduced
radio range of ground-to-ground transmissions, and because the rescue
workers move quite slowly, the replicas maintained in their handheld
device do not have many opportunities to synchronize directly with one
another (only 7.6\% of the contacts occur between two replicas, see
Table~\ref{fig:disaster_syncs}). They depend on the drones to
disseminate information over long distances. A snapshot of the
simulation area is shown in Fig.~\ref{fig:disaster_snapshot}. This
figure shows only part of the $10\times\!10\,$ km${{}^2}$ area.

\begin{table}[h]
\centering{}\caption{\label{fig:disaster_details}Details about the Disaster Relief scenario}
\begin{tabular}{|c|c|}
\cline{2-2} 
\multicolumn{1}{c|}{} & {\small{}\hspace*{-1ex}\rule{0pt}{2.5ex}\textbf{Values ({*} = min/max/avg/stdev)}\hspace*{-1ex}}\tabularnewline
\hline
{\small{}\hspace*{-1ex}\rule{0pt}{2.5ex}}\textbf{\small{}A: Mobility
scenario for replicas}{\small{}\hspace*{-1ex}} & {\small{}Levy Walk~\cite{ton11rhee}}\tabularnewline
\hline 
{\small{}\rule{0pt}{2.5ex}Duration of the scenario} & {\small{}24 hours}\tabularnewline
\hline 
{\small{}\rule{0pt}{2.5ex}Nb of nodes} & {\small{}1000}\tabularnewline
\hline 
{\small{}\rule{0pt}{2.5ex}Speed} & {\small{}{[}1..2{]} m/s}\tabularnewline
\hline 
{\small{}\rule{0pt}{2.5ex}Delay between flights} & {\small{}{[}0..10{]} s}\tabularnewline
\hline 
{\small{}\rule{0pt}{2.5ex}Flight distances} & {\small{}{[}0..100{]} m}\tabularnewline
\hline 
{\small{}\rule{0pt}{2.5ex}Levy Walk~specific parameters} & {\small{}$\alpha=1.8\,\,\,\,\,\beta=\alpha$$\,\,\,\,\,\,k=30.55\,\,\,\,\,\rho=0.89$}\tabularnewline
\hline 
\hline
\textbf{\small{}B: Mobility scenario for relays}{\small{}\hspace*{-1ex}} & {\small{}Random Waypoint}\tabularnewline
\hline 
{\small{}\rule{0pt}{2.5ex}Duration of the scenario} & {\small{}24 hours}\tabularnewline
\hline 
{\small{}\rule{0pt}{2.5ex}Nb of nodes} & {\small{}100}\tabularnewline
\hline 
{\small{}\rule{0pt}{2.5ex}Speed} & {\small{}{[}5..20{]} m/s}\tabularnewline
\hline 
{\small{}\rule{0pt}{2.5ex}Flight distances} & {\small{}{[}0..15000{]} m}\tabularnewline
\hline
\hline
\textbf{\small{}Contact scenario} & {\small{}\rule{0pt}{2.5ex}Based on a combination of A and B}\tabularnewline
\hline 
{\small{}\rule{0pt}{2.5ex}Transmission range} & {\small{}50 m ground-to-ground, 200 m air-to-ground or air-to-air}\tabularnewline
\hline 
{\small{}\rule{0pt}{2.5ex}Number of contacts between pairs of nodes} & {\small{}439552}\tabularnewline
\hline 
{\small{}\rule{0pt}{2.5ex}Duration of contacts} & {\small{}$5''$ / $15'35''$ / $34''$ / $33''$ }\textsuperscript{{\small{}({*})}}\tabularnewline
\hline 
\hline 
\textbf{\small{}Application scenario} & {\small{}\rule{0pt}{2.5ex}258000 update operations between 00:05 and
21:36}\tabularnewline
\hline 
\end{tabular}
\end{table}

\subsubsection{Application scenario}

A simple application scenario is defined, whereby each node carrying a
replica (i.e., a rescue worker using a handheld device) issues an
update operation on its local state once every 5 minutes in the
{[}00:05{\small{}--}21:35{]} interval, with a randomly selected
jitter at start time so that all replicas behave
asynchronously. Overall, 258000 update operations are issued on the 1000
replicas (258 each) during the interval
considered.

\subsubsection{Simulation results}

\begin{figure}[H]
\begin{centering}
\includegraphics[width=17.5cm]{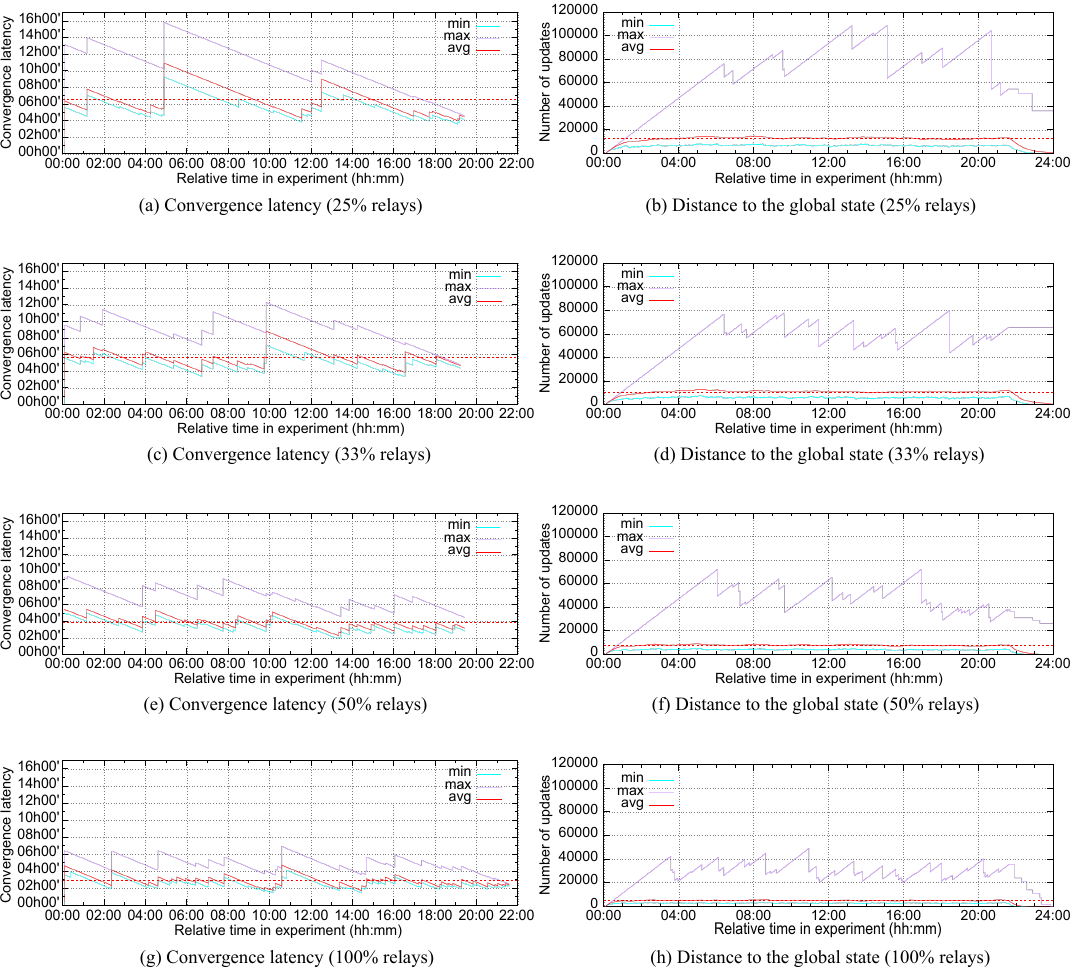}
\end{centering}
\caption{\label{fig:disaster1}Evolution of the convergence
latency (left) and distance to the global state (right) when running
the Disaster relief scenario with different ratios of relays}
\end{figure}

\begin{figure}[H]
\begin{centering}
\includegraphics[width=17.5cm]{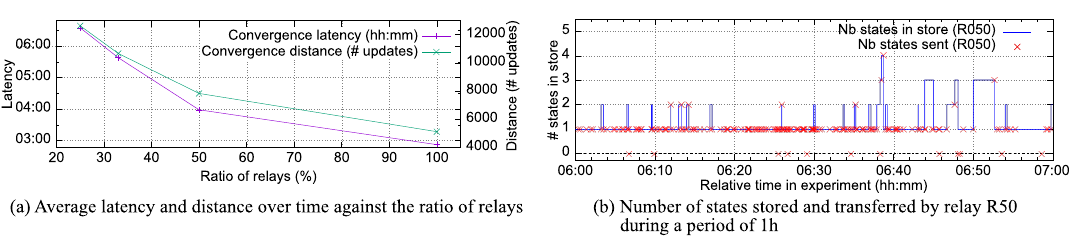}
\end{centering}
\caption{\label{fig:disaster2}Evolution of the average latency
and distance over time against the ratio of relays (left), and evolution
of the number of states stored and transferred by a single relay (right)
while running the Disaster relief scenario}
\end{figure}

Fig.~\ref{fig:disaster1} and~\ref{fig:disaster2} present results that
have been obtained with the Disaster Relief scenario, using the same
experimental procedure and the same metrics as for the \"Ostermalm and
VBN scenarios, with different ratios of drones serving as airborne
relays (from 25\% to 100\%). Tables~\ref{fig:disaster_syncs}
to~\ref{fig:disaster_replica_tsfers} show statistics on the number of
synchronizations and the number of states transferred between relays
and replicas.

In Fig.~\ref{fig:disaster1} it can be observed that when only 25\% of
the drones serve as relays, the average latency over time is around
$6\mathrm{h}34'$, with an average distance of 12630 update operations.
Moreover, the convergence latency shown in
Fig.~\ref{fig:disaster1}{\textcolor{blue}a} is not defined after
19:25, which indicates that the global state reached after that time
will never be seen by any replica. This is consistent with the fact
that the average latency at 19:25 is about $4\mathrm{h}26'$: any value
of the global state reached after 19:25 could only be perceived by
most replicas after the end of the $24\mathrm{h}$ simulation period
considered in this scenario.

When the number of drones serving as relays is increased, the
convergence latency and distance decrease significantly, and reach
respectively $2\mathrm{h}52'$ and 5122 update operations on average
when 100\% of the drones are used as relays. In the latter case, all
the replicas converge at the end of the simulation period, as shown in
Fig.~\ref{fig:disaster1}{\textcolor{blue}h}.

Even when 50\% or 33\% of the drones serve as relays, the synchronization of
the replicas is significantly improved compared with a scenario involving only
25\% relays. Some of the replicas fail to see the last updates, though, as
shown in Fig.~\ref{fig:disaster2}{\textcolor{blue}a} and
Fig.~\ref{fig:disaster2}{\textcolor{blue}b}.

Fig.~\ref{fig:disaster2}{\textcolor{blue}b} shows the evolution of the
number of states maintained by relay R050 in its local store between
06:00 and 07:00, and the number of states transferred by this relay
when synchronizing with neighbor nodes during that interval. Since
1000 replicas are involved in this scenario, each relay's store may
theoretically accumulate up to 1000 conflicting states. Yet it can be
observed that R050's store only contains a couple of states at any
time. This is representative of what occurs on any relay during the
whole simulation interval, as confirmed in
Tables~\ref{fig:disaster_relay_stored} and
~\ref{fig:disaster_relay_tsfers}.

The fact that a large number of replicas does not imply that the
relays have to maintain many conflicting states in their store is
interesting, for it has a direct impact on the cost and scalability of
relay-assisted synchronization. This is discussed further in
Section~\ref{sec:Discussion}.

\begin{table*}[h]
\centering{}\caption{\label{fig:disaster_syncs}Statistics about the number of synchronizations
observed against the ratio of relays (Disaster relief scenario)}
{\small{}}%
\begin{tabular}{|c|c|c|c|c|}
\hline
{\small Ratio of drones serving as relays} & {\small 25\% (25 drones)} & {\small 33\% (33 drones)} & {\small 50\% (50 drones)} & {\small 100\% (100 drones)}\tabularnewline
\hline
\hline
{\small Nb. of sync. between pairs of nodes} & {\small 110839} & {\small 164707} & {\small 229533} & {\small 439552}\tabularnewline
\hline
{\small Nb. of sync. between pairs of replicas} & {\small 34831} & {\small 34831} & {\small 34831} & {\small 34831}\tabularnewline
\hline
{\small Nb. of sync. between replicas and relays} & {\small 74670} & {\small 126021} & {\small 186320} & {\small 371164}\tabularnewline
\hline
{\small Nb. of sync. between pairs of relays} & {\small 1338} & {\small 3855} & {\small 8382} & {\small 33557}\tabularnewline
\hline
\end{tabular}
\end{table*}

\begin{table*}[h]
\centering{}\caption{\label{fig:disaster_relay_stored}Statistics about the number of
states stored by relays during synchronizations (Disaster relief scenario)}
{\small{}}%
\begin{tabular}{|c|c|c|c|c|c|c|c|c|c|c|}
\cline{2-11}
\multicolumn{1}{c|}{} & {\small Nb. of states} & \multirow{2}{*}{{\small 0}} & \multirow{2}{*}{{\small 1}} & \multirow{2}{*}{{\small 2}} & \multirow{2}{*}{{\small 3}} & \multirow{2}{*}{{\small 4}} & \multirow{2}{*}{{\small 5}} & \multirow{2}{*}{{\small 6}} & \multirow{2}{*}{{\small 7}} & \multirow{2}{*}{{\small Total}}\tabularnewline
\multicolumn{1}{c|}{} & {\small in stores} &  &  &  &  &  &  &  &  & \tabularnewline
\hline
{\small 100\%} & {\small Nb. of occurrences} & {\small 14} & {\small 440308} & {\small 37004} & {\small 6283} & {\small 1139} & {\small 224} & {\small 48} & {\small 8} & {\small 485208}\tabularnewline
\cline{2-11}
{\small relays} & {\small Ratio} & {\small 0.003\%} & {\small 90.7\%} & {\small 7.6\%} & {\small 1.2\%} & {\small 0.2\%} & {\small 0.05\%} & {\small 0.01\%} & {\small 0.001\%} & {\small 100\%}\tabularnewline
\hline
\hline
{\small 50\%} & {\small Nb. of occurrences} & {\small 8} & {\small 202726} & {\small 9730} & {\small 868} & {\small 77} & {\small 14} & {\small -} & {\small -} & {\small 213423}\tabularnewline
\cline{2-11}
{\small relays} & {\small Ratio} & {\small 0.004\%} & {\small 94.9\%} & {\small 4.5\%} & {\small 0.4\%} & {\small 0.04\%} & {\small 0.006\%} & {\small -} & {\small -} & {\small 100\%}\tabularnewline
\hline
\end{tabular}
\end{table*}
\begin{table*}[h]
\centering{}\caption{\label{fig:disaster_relay_tsfers}Statistics about the number of
states transferred by relays during synchronizations (Disaster relief scenario)}
{\small{}}%
\begin{tabular}{|c|c|c|c|c|c|c|c|c|c|}
\cline{2-10}
\multicolumn{1}{c|}{} & {\small Nb. of states transferred} & \multirow{2}{*}{{\small 0}} & \multirow{2}{*}{{\small 1}} & \multirow{2}{*}{{\small 2}} & \multirow{2}{*}{{\small 3}} & \multirow{2}{*}{{\small 4}} & \multirow{2}{*}{{\small 5}} & \multirow{2}{*}{{\small 6}} & \multirow{2}{*}{{\small Total}}\tabularnewline
\multicolumn{1}{c|}{} & {\small per sync.} &  &  &  &  &  &  &  & \tabularnewline
\hline
{\small 100\%} & {\small Nb. of occurrences} & {\small 30356} & {\small 367714} & {\small 17601} & {\small 1916} & {\small 281} & {\small 40} & {\small 6} & {\small 417914}\tabularnewline
\cline{2-10}
{\small relays} & {\small Ratio} & {\small 7.2\%} & {\small 87.9\%} & {\small 4.2\%} & {\small 0.4\%} & {\small 0.07\%} & {\small 0.009\%} & {\small 0.001\%} & {\small 100\%}\tabularnewline
\hline
\hline
{\small 50\%} & {\small Nb. of occurrences} & {\small 7567} & {\small 183575} & {\small 5177} & {\small 310} & {\small 28} & {\small 2} & {\small -} & {\small 196659}\tabularnewline
\cline{2-10}
{\small relays} & {\small Ratio} & {\small 38\%} & {\small 93.3\%} & {\small 26\%} & {\small 0.1\%} & {\small 0.01\%} & {\small 0.001\%} & {\small -} & {\small 100\%}\tabularnewline
\hline
\end{tabular}
\end{table*}

\begin{table*}[h]
\centering{}\caption{\label{fig:disaster_replica_tsfers}Statistics about the number of
states transferred by replicas during synchronizations (Disaster relief
scenario)}
{\small{}}%
\begin{tabular}{|c|c|c|c|c|}
\cline{2-5}
\multicolumn{1}{c|}{} & {\small Nb. of states transferred per sync.} & {\small 0} & {\small 1} & {\small Total}\tabularnewline
\hline
\multirow{2}{*}{{\small 100\% relays}} & {\small Nb. of occurrences} & {\small 27936} & {\small 392526} & {\small 420462}\tabularnewline
\cline{2-5}
 & {\small Ratio} & {\small 6.6\%} & {\small 93.4\%} & {\small 100\%}\tabularnewline
\hline
\hline
\multirow{2}{*}{{\small 50\% relays}} & {\small Nb. of occurrences} & {\small 23646} & {\small 225911} & {\small 249557}\tabularnewline
\cline{2-5}
 & {\small Ratio} & {\small 9.5\%} & {\small 90.5\%} & {\small 100\%}\tabularnewline
\hline
\end{tabular}
\end{table*}


\section{Discussion\label{sec:Discussion}}

The simulation results presented in the former section confirm that
using relays can help the replicas to converge in a variety of situations.
In this section we examine the scalability of this approach, its impact
on the security of the data shared by the replicas, and its usability
outside of OppNets.

\subsection{Scalability\label{subsec:Scalability}}
The synchronization protocols defined in
Section~\ref{sec:Synchronization_algorithms} use version vectors to
characterize the causal context of each replica's state. These version
vectors can be implemented as maps of (key, counter) pairs, where each
key is the identifier of a replica. The weight of version vectors
therefore depends solely on the number of replicas involved in a
synchronization scenario. It is independent from the number of relays
that assist in this synchronization.

When two nodes synchronize, the number of states they exchange is kept
at a minimum. More specifically, when a relay synchronizes with
another node (either replica or relay), only a subset of the states
contained in its store ---those that have been selected to inflate the
peer's state--- are transferred to the peer node.

Although the number of conflicting states maintained in each relay's
store may theoretically be as high as the number of replicas that
issue such states (for a given CRDT instance), the experiments whose
results are presented in Section~\ref{sec:Experimentation} show that
the stores usually contain a far smaller number of states, even if the
number of replicas is high.

The fact that a relay can hardly accumulate many conflicting states in
its store is interesting, as this has a direct impact on the overall
cost of relay-assisted synchronization. Indeed, since a relay ends up
with a single state in its store after each synchronization with a
replica, it can only gain additional conflicting states by
synchronizing with other relays \emph{before synchronizing again with
a replica}. When the contact scenario is such that a relay meets
replicas frequently, its store gets purged regularly. This is
typically the case in the Disaster Relief Scenario considered in
Section~\ref{subsec:Scenario_Disaster}: this scenario involves 1000
replicas, but after leaving a replica each relay (drone) can only
accumulate a couple of states in its store before synchronizing again
with a replica. As a consequence, the computational complexity of the
selection process (for a relay) remains reasonable, and so does the
message complexity, which is mostly driven by the number of states
transferred during synchronizations (assuming states weigh a lot more
than version vectors and trigger messages).

It would certainly be possible to devise an adverse scenario involving
a large population of replicas (so each relay's store can potentially
accumulate many conflicting states), a large population of relays (so
each relay gets many chances to synchronize with other relays), with
contact patterns such that each relay can meet many other relays
between two successive meetings with replicas. Yet, while devising the
experimentation scenarios presented in this paper we failed to
identify a plausible scenario bearing such unfavorable
characteristics. Thus, although a relay may occasionally maintain a
large number of conflicting states in its store (possibly close to the
number of replicas), we believe that in most realistic use-cases the
number of states carried by relays will be far smaller than the number
of replicas they assist.
  
\subsection{Security\label{subsec:Security}}
It has been shown in~\cite{sicherheit22jacob} that state-based CRDTs
are inherently crash-tolerant and Byzantine-tolerant, as long as eventual
delivery is ensured between correct replicas. Yet the given demonstration
is based on the assumption that end-to-end connectivity is ensured
between all replicas, and that these replicas synchronize pairwise
via authenticated channels. In the present paper we consider that
replicas can synchronize either directly with one another, or via
relays. The question whether the relays may somehow disrupt the synchronization
of the replicas must therefore be considered.

As suggested in Section~\ref{sec:System_model}, the fact that relays
need only to perceive the replicas' states they process as simple
blobs of opaque data makes it possible for the replicas to encrypt
and sign every state they transfer to a relay, and to verify the signature
of every state they receive from a replica or relay. This requires
either that all replicas share a common secret key, or that each replica
holds an asymmetric key pair (i.e., private and public key), as well
as a key store of all the other replicas' public keys. Thus, when
receiving a state from a relay, a replica can check that this state
has been issued by a legitimate replica. A Byzantine relay could not
deceive a replica by issuing fake states or by altering the states
they got from legitimate replicas. Besides, since the merge function
in a state-based CRDT is commutative, associative and idempotent,
a Byzantine relay could also not disrupt the synchronization of replicas
by issuing the same replica's state several times, or by changing
the order in which states are delivered to replicas. 

Although a Byzantine relay cannot disrupt the replicas directly, it
may however disrupt the behavior of other relays by providing them
with fake states associated with fake version vectors, thus inciting
these relays to purge their store from valid states and replace them
with invalid states. Although these states would eventually be discarded
by the replicas, this kind of attack would however hamper the propagation
of valid states, thus reducing the chance that these valid states
eventually reach the replicas. In a scenario where the replicas can
still synchronize directly with one another (such as in the examples
depicted in Sections~\ref{subsec:Scenario_Ostermalm} and~\ref{subsec:Scenario_Disaster}),
this kind of attack would simply increase the convergence latency
and convergence distance of each replica. When the replicas are fully
dependent on the relays to synchronize (as depicted in Section~\ref{subsec:Scenario_VBN}),
a small population of Byzantine relays may indeed create a permanent
partition of the replicas, by preventing valid states to reach all
replicas. This would in effect contradict the SEC condition, which
states that all updates must be eventually delivered to ---and applied
on--- all replicas. This kind of attack may be prevented by having
each relay check the validity (signature) of each state received from
another relay before inserting this state in its own store, but this
would require that all relays hold the public key ---or keys---
of the replicas. Depending on the kind of deployment scenario considered,
such an approach may be acceptable or not. For example, in the scenario
considered in Section~\ref{subsec:Scenario_VBN}, each relay carried
by a bus may hold the public keys of the replicas located in public
service buildings. In contrast, in the scenario considered in Section~\ref{subsec:Scenario_Ostermalm},
each relay device carried by a pedestrian entering the \"Ostermalm district
may hardly hold the public keys of the replicas roaming this district.
In some deployment scenarios, an approach based on federated learning
may be used to allow the replicas to learn which relays they can trust,
and to propagate this information among replicas and among trusted
relays. 

Further considerations about this topic fall outside the scope of
this paper and are left for future work.

\subsection{From relays in OppNets to relays in the cloud\label{subsec:Cloud}}
In this paper we have focused on the idea of deploying relays in OppNets
in order to facilitate the synchronization of replicas. In such networks
direct synchronization between replicas is hampered by the scarcity
of radio contacts between these replicas. Relays can improve the propagation
of replica's states, thus speeding up the synchronization process,
and ultimately the convergence of all replicas.

Yet the idea of using relays for replica synchronization also makes
sense for more traditional networking environments, such as the Internet.
Relays may for example be deployed in datacenters (in the cloud),
and thus serve as stable repositories for client-side replicas and
applications. Most of the current use cases of CRDTs rely on this
typical client-server model, albeit with dedicated ``relays'' on
the server-side: clients designed with the Yjs library require servers
designed with the Yjs library, and the same goes for clients based
on Automerge, AntidoteDB, etc. 

With the synchronization protocols we have defined, the relays only
perceive replicas's states as encoded byte arrays. A single relay
may thus be deployed on a server, and assist in the synchronization
of many replicas, regardless of the CRDT libraries used to implement
these clients. Moreover, as explained above the states issued by these
replicas may be encrypted and signed, thus preventing any attempt
to alter valid states or forge invalid ones, on the server's side
or in the network.

\section{\label{sec:Conclusion}Conclusion}

In this paper we have investigated the idea of using relays to speed
up the convergence of state-based CRDT replicas in an opportunistic
network (OppNet), using only transient radio contacts between two
nodes (either replicas or relays) to ensure the synchronization of
these nodes. In the model we consider, some of the nodes in an OppNet
can behave as relays in order to assist those nodes that host replicas
to synchronize their local states. The protocols we designed to ensure
this relay-assisted synchronization do not require that the relays
have any knowledge about the data they are carrying. In essence, every
relay perceives the state issued by a replica as a simple blob of
opaque data, which may even be signed and encrypted by the issuer,
so that only legitimate replicas can process the states they receive
from relays.

Simulations have been conducted in order to show how relay-assisted
synchronization can be achieved in different scenarios, depending on
the mobility patterns of replicas and relays. Novel metrics pertaining
to convergence latency and convergence distance have been defined in
order to quantify the gain observed when using relays. The results
confirm that the approach is scalable, that it shows reasonable time
complexity and message complexity, and that using relays can
significantly reduce the convergence latency and convergence distance
observed on replicas.

Besides being used to produce simulation results, the synchronization
protocols presented in this paper have also being implemented in a
demonstrator we developed to illustrate how CRDT-based applications
can be used in real opportunistic networks. The code of this
demonstrator is available on our Web
site~\footnote{\href{https://www-inzu.irisa.fr/demo-oppnet}{https://www-inzu.irisa.fr/demo-oppnet}}.

\bibliography{refs}

@InBook{appalgo97hochbaum,
  author = {Hochbaum, Dorit S.},
  title = {Approximation Algorithms for NP-Hard Problems},
  chapter = {3, Approximating Covering and Packing Problems: Set Cover, Vertex Cover, Independent Set And Related Problems},
  publisher = {PWS Publishing Company},
  year = 1996,
  pages = {94-143}
}

@Inproceedings{sigcomm03fall,
 author = {Kevin Fall},
 title = {{A Delay-Tolerant Network Architecture for Challenged Internets}},
 booktitle = {ACM Annual Conference of the Special Interest Group on Data Communication (SIGCOMM 2003)},
 address = {Karlsruhe, Germany},
 pages = {27-34},
 year = {2003},
 month = aug,
 doi = {10.1145/863955.863960}
}

@Inproceedings{sicherheit22jacob,
 title = {{On CRDTs in Byzantine Environments}},
 author = {Jacob, Florian and Bayreuther, Saskia and Hartenstein, Hannes},
 booktitle = {Sicherheit 2022},
 address = {Karlsruhe, Germany},
 year = {2022},
 month = apr,
 pages = {113-122},
 publisher = {Gesellschaft f\"ur Informatik},
 doi = {10.18420/sicherheit2022_07}
}

@Misc{kth/walkers,
 author = {Kouyoumdjieva, Sylvia Todorova and Helgason, \'{O}lafur Ragnar and Karlsson, Gunnar},
 howpublished = {IEEE Dataport},
 title = {CRAWDAD kth/walkers},
 year = {2022},
 doi = {10.15783/C7Z30C}
}

@Misc{ubs/vbn,
 author = {Guidec, Fr\'{e}d\'{e}ric and Launay, Pascale and Mah\'{e}o, Yves},
 howpublished = {IEEE Dataport},
 title = {CRAWDAD ubs/vbn},
 year = {2022},
 doi = {10.15783/qr0f-m304}
}

@Inproceedings{sapir04lindgren,
 author = {Lindgren, Anders and Doria, Avri and Schelen, Olov},
 title = {{Probabilistic Routing in Intermittently Connected Networks}},
 booktitle = {1st International Workshop on Service Assurance with Partial and Intermittent Resources (SAPIR 2004)},
 year = {2004},
 month = aug,
 address = {Fortaleza, Brazil},
 series = {LNCS},
 volume = {3126},
 pages = {239-254},
 publisher = {Springer},
 doi = {10.1007/978-3-540-27767-5_24}
}

@InProceedings{coordination09imine,
author = {Imine, Abdessamad},
title = {{Coordination Model for Real-Time Collaborative Editors}},
booktitle = {Coordination Models and Language (COORDINATION 2009)},
address = {Lisbon, Portugal},
year = 2009,
month = jun,
volume = {5521},
series = {LNCS} ,
publisher = {Springer},
pages = {225-246},
doi = {10.1007/978-3-642-02053-7_12}
}

@Inproceedings{icta17alsulami,
 author = {Alsulami, Noha and Cherif, Asma and Imine, Abdessamad},
 booktitle = {6th International Conference on Information and Communication Technology and Accessibility (ICTA)}, 
 title = {{Evaluating Data Convergence of Collaborative Editors in Opportunistic Networks}}, 
 year = {2017},
 month = dec,
 address = {Muscat, Oman},
 publisher = {IEEE},
 doi = {10.1109/ICTA.2017.8336061}
}

@Article{computingsurveys24almeida,
 author = {Almeida, Paulo S\'{e}rgio},
 title = {{Approaches to Conflict-free Replicated Data Types}},
 year = {2024},
 volume = {57},
 number = {1},
 pages = {1-36},
 publisher = {ACM},
 journal = {ACM Computing Surveys},
 month = sep,
 doi = {10.1145/3695249}
}

@Article{ieeecomst15chakchouk,
 author = {Chakchouk, Nessrine},
 title = {{A Survey on Opportunistic Routing in Wireless Communication Networks}},
 journal = {{IEEE Communications Surveys \& Tutorials}},
 year = 2015,
 volume = 17,
 number = 4,
 pages = {2214-2241},
 doi = {10.1109/COMST.2015.2411335}
}

@Article{ieeeaccess22dalal,
 author = {Dalal, Renu and Khari, Manju and Anzola, John Petearson and Garc\'ia-D\'iaz, Vicente},
 title = {{Proliferation of Opportunistic Routing: A Systematic Review}}, 
 journal = {IEEE Access},
 volume = {10},
 year = {2022},
 pages = {5855-5883},
 doi = {10.1109/ACCESS.2021.3136927}
}

@Misc{arxiv13almeida,
 author = {Almeida, Paulo S\'{e}rgio and Baquero, Carlos},
 title = {{Scalable Eventually Consistent Counters over Unreliable Networks}},
 year = {2013},
 month = jul,
 howpublished = {arXiv 1307.3207},
 doi = {10.48550/arXiv.1307.3207}
}

@Inproceedings{icoin18robin,
 author = {Robin, Charles Edward A. and Romero, Victor M.},
 title = {{DTNDocs: A delay tolerant peer-to-peer collaborative editing system}},
 booktitle = {32nd International Conference on Information Networking (ICOIN)},
 address = {Chiang Mai, Thailand},
 year = {2018},
 month = jan,
 pages = {92-97},
 doi = {10.1109/ICOIN.2018.8343092}
}

@Inproceedings{noms14ciobanu,
 author = {Ciobanu, Radu-Ioan and Marin, Radu-Corneliu and Dobre, Ciprian and Cristea, Valentin and Mavromoustakis, Constandinos X.},
 title = {{ONSIDE: Socially-aware and Interest-based Dissemination in Opportunistic Networks}},
 booktitle = {IEEE Network Operations and Management Symposium (NOMS)},
 address = {Krakow, Poland},
 month = may,
 year = {2014},
 publisher = {IEEE},
 doi = {10.1109/NOMS.2014.6838390}
}

@Inproceedings{icdcs09weiss,
 author = {Weiss, Stephane and Urso, Pascal and Molli, Pascal},
 title = {{Logoot: A Scalable Optimistic Replication Algorithm for Collaborative Editing on P2P Networks}},
 booktitle = {29th IEEE International Conference on Distributed Computing Systems (ICDCS'09)},
 address = {Montreal, Canada},
 year = {2009},
 month = jun,
 pages = {404-412},
 publisher = {IEEE},
 doi = {10.1109/ICDCS.2009.75}
}

@TechReport{inria11shapiro,
 author = {Shapiro, Marc and Pregui{\c c}a, Nuno and Baquero, Carlos and Zawirski, Marek},
 title = {{A Comprehensive Study of Convergent and Commutative Replicated Data Types}},
 institution =  {INRIA},
 number = {7506},
 year = {2011},
 month = jan
}

@Inproceedings{chants14karkkainen,
 author = {K\"{a}rkk\"{a}inen, Teemu and Ott, J\"{o}rg},
 title = {{Shared Content Editing in Opportunistic Networks}},
 year = {2014},
 publisher = {ACM},
 booktitle = {9th ACM MobiCom Workshop on Challenged Networks (CHANTS'14)},
 pages = {61--64},
 doi = {10.1145/2645672.2645685}
}

@Misc{arxiv18preguica,
 title = {{Conflict-free Replicated Data Types: an Overview}},
 author = {Pregui\c{c}a, Nuno},
 year = {2018},
 month = jun,
 howpublished = {arXiv 1806.10254},
 numpages = {4},
 doi = {10.48550/arXiv.1806.10254}
}

@Article{wowkivs11schildt,
 author = {Sebastian Schildt and Johannes Morgenroth and Wolf-Bastian P{\"o}ttner and Lars Wolf},
 journal = {Electronic Communications of the EASST},
 title = {{IBR-DTN: A lightweight, modular and highly portable Bundle Protocol implementation}},
 pages = {1-11},
 volume = {37},
 year = {2011},
 month = jan
}

@Inbook{igi16costea,
 title = {{Causal and Total Order in Opportunistic Networks}},
 author = {Costea, Mihail and Ciobanu, Radu-Ioan and Marin, Radu-Corneliu and Dobre, Ciprian and Mavromoustakis, Constandinos X. and Mastorakis, George},
 chapter = {Emerging Innovations in Wireless Networks and Broadband Technologies},
 pages = {221-262},
 year = {2016},
 publisher = {IGI Global},
 doi = {10.4018/978-1-4666-9941-0.ch010}
}

@Article{ccpe17costea,
 title = {{Total Order in Opportunistic Networks}},
 author = {Costea, Mihail and Ciobanu, Radu-Ioan and Marin, Radu-Corneliu and Dobre, Ciprian and Mavromoustakis, Constandinos X. and Mastorakis, George and Xhafa, Fatos},
 year = {2017},
 journal = {Concurrency and Computation: Practice and Experience},
 volume = {29},
 number = {10},
 publisher = {Wiley Online Library},
 doi = {10.1002/cpe.4056}
}

@Article{comcom16pajevic,
 author = {Pajevic, Ljubica and Karlsson, Gunnar},
 title = {{Modeling Opportunistic Communication with Churn}},
 journal = {Computer Communications},
 volume = {96},
 pages = {123-135},
 year = 2016,
 issn = {0140-3664},
 doi = {10.1016/j.comcom.2016.04.018}
}

@Article{ton11rhee,
 author = {Rhee, Injong and Shin, Minsu and Hong, Seongik and Lee, Kyunghan and Kim, Seong Joon and Chong, Song},
 title = {{On the Levy-walk Nature of Human Mobility}},
 journal = {IEEE/ACM Transactions on Networking},
 volume = {19},
 number = {3},
 month = jun,
 year = {2011},
 pages = {630-643},
 publisher = {IEEE},
 doi = {10.1109/TNET.2011.2120618}
}

@Techreport{tr00vahdat,
 author = {Vahdat, Amin and Becker, David},
 title = {{Epidemic Routing for Partially Connected Ad Hoc Networks}},
 institution = {Duke University, Durham, USA},
 year = {2000},
 month = apr,
 number = {CS-200006}
}

@InProceedings{ubicomm23maheo,
 author = {Guidec, Fr\'ed\'eric and Mah\'eo, Yves and No\^us, Camille},
 title = {{CRDT-based Collaborative Editing in OppNets: a Practical Experiment}},
 booktitle = {17th Conference on Mobile Ubiquitous Computing, Systems, Services and Technologies (Ubicomm 2023)},
 year = 2023,
 pages = {13-21},
 month = sep,
 address = {Porto, Portugal},
 publisher = {IARIA}
}

@Article{ppna22guidec,
 author = {Guidec, Fr\'ed\'eric and Mah\'eo, Yves and No\^us, Camille},
 title = {{Supporting conflict-free replicated data types in opportunistic networks}},
 journal = {Peer-to-Peer Networking and Applications},
 month = jan,
 year = {2023},
 volume = {16},
 pages = {395-419},
 publisher = {Springer},
 doi = {10.1007/s12083-022-01404-6}
}
\bibliographystyle{plainurl}

\clearpage

\appendix

\section*{Appendix A\quad Enhanced synchronization protocol\label{sec:enhanced_algorithms}}

\noindent NOTA: In the following algorithms, the assumption is made that the execution of event handlers is atomic, as in Javascript.

\begin{algorithm}[h]
  \includegraphics[width=9.5cm]{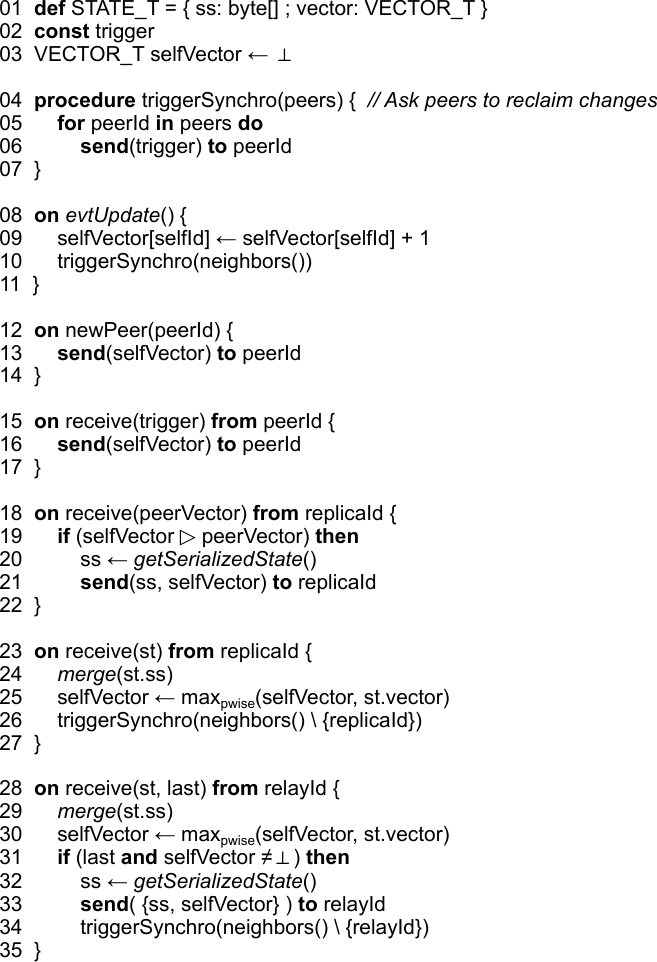}\par 
  \caption{\label{alg:full-replica-algo}Enhanced synchronization protocol (replica side)}
\end{algorithm}

\begin{algorithm}[h]
  \includegraphics[width=10.5cm]{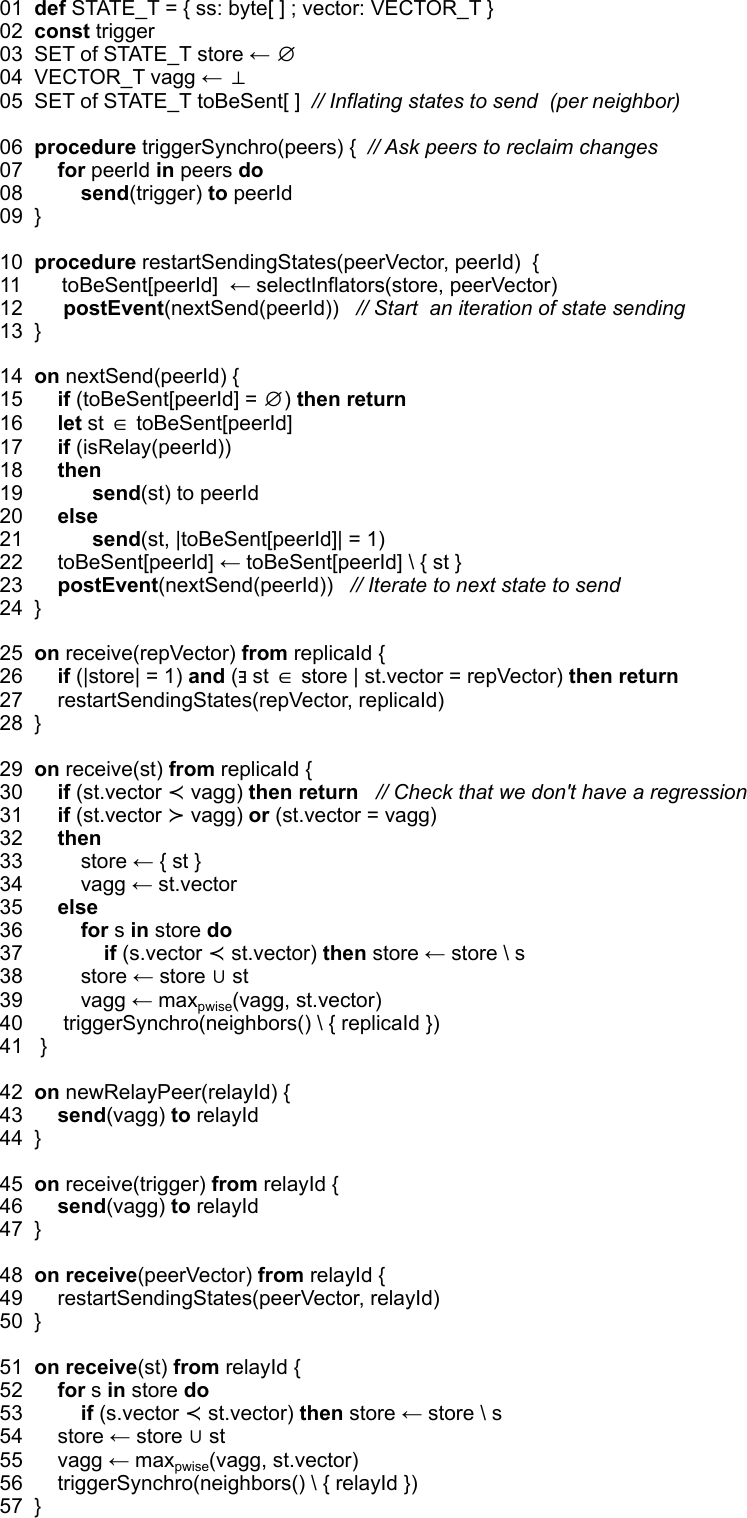}\par 
  \caption{\label{alg:full-relay-algo}Enhanced synchronization protocol (relay side)}
\end{algorithm}

\clearpage

\end{document}